\documentclass[aps,twocolumn,pra,tightenlines,floatfix,showpacs,superscriptaddress]{revtex4-1}
\usepackage{graphicx}
\usepackage{amsmath}
\usepackage{amssymb}
\usepackage{times}
\usepackage[english]{babel}

\begin{document}
\title{Thermodynamics and structural transition of binary atomic Bose-Fermi mixtures in box or harmonic potentials: A path-integral study}

\author{Tom Kim}
\affiliation{School of Natural Sciences, University of California, Merced, CA 95343, USA}

\author{Chih-Chun Chien}
\affiliation{School of Natural Sciences, University of California, Merced, CA 95343, USA}
\email{cchien5@ucmerced.edu}

\begin{abstract}
Experimental realizations of a variety of atomic binary Bose-Fermi mixtures have brought opportunities for studying composite quantum systems with different spin-statistics. The binary atomic mixtures can exhibit a structural transition from a mixture into phase separation as the boson-fermion interaction increases. By using a path-integral formalism to evaluate the grand partition function and thermodynamic grand potential, we obtain the effective potential of binary Bose-Fermi mixtures. Thermodynamic quantities in a broad range of temperatures and interactions are also derived. The structural transition can be identified as a loop of the effective potential curve, and the volume fraction of phase separation can be determined by the lever rule. For $^6$Li-$^7$Li and $^6$Li-$^{41}$K mixtures, we present the phase diagrams of the mixtures in a box potential at zero and finite temperatures. Due to the flexible densities of atomic gases, the construction of phase separation is more complicated when compared to conventional liquid or solid mixtures where the individual densities are fixed. For harmonically trapped mixtures, we use the local density approximation to map out the finite-temperature density profiles and present typical trap structures, including the mixture, partially separated phases, and fully separated phases.
\end{abstract}


\maketitle

\section{Introduction}
The study of binary atomic Bose-Fermi mixtures has a long history. ${}^3$He - ${}^4$He mixtures exhibit phase separation between a $^3$He-rich phase and a $^3$He-poor phase at low temperatures~\cite{10.2307/1744198,Kuerten1987,pobell2013matter}. Pumping $^3$He across the separation is the main mechanism of dilution refrigerators~\cite{Lounasmaa_book}. After the observation of Bose-Einstein condensation (BEC) in ultracold atoms~(see Refs.~\cite{RevModPhys.71.463,pethick2008bose,ueda2010fundamentals} for a review), further advancements have made it possible to mix ultracold bosons and fermions with different spin-statistics. For example, binary atomic ${^6}$Li and ${^7}$Li mixtures was achieved in 2001~\cite{PhysRevLett.87.080403,Truscott2570}. Then, other binary atomic Bose-Fermi mixtures have been produced, including ${}^{40}$K and ${}^{87}$Rb~\cite{PhysRevLett.89.150403}, ${}^{6}$Li and ${}^{87}$Rb~\cite{PhysRevA.77.010701},  ${}^{84}$Sr and ${}^{87}$Sr~\cite{PhysRevA.82.011608},  ${}^6$Li and ${}^{41}$K~\cite{PhysRevLett.117.145301,Lous18}, ${}^6$Li and ${}^{133}$Cs~\cite{PhysRevLett.119.233401}, etc. A review on trapping and cooling atomic Bose-Fermi mixtures can be found in Ref.~\cite{1063-7869-59-11-1129}. Interestingly, the different behavior of the specific heat of bosons and fermions can limit sympathetic cooling of ultracold atoms~\cite{Presilla03}. 
Although the tunable atom-atom interactions can be either attractive or repulsive~\cite{pethick2008bose,1063-7869-59-11-1129}, attractive bosons are unstable against collapse at low temperatures~\cite{Donley2001} and may lead to further complications. Hence, here we will focus on binary atomic Bose-Fermi mixtures with all repulsive interactions. 

On the theoretical side, binary atomic Bose-Fermi mixtures with repulsive interactions at zero temperature in a harmonic trap have been studied~\cite{PhysRevLett.80.1804,PhysRevA.59.2974}, and the systems are shown to exhibit a structural transition into phase separation if the inter-species repulsion is too strong. The stability conditions for uniform mixtures have been studied in three-dimension ~\cite{PhysRevA.61.053605} and in mixed dimensions~\cite{PhysRevA.88.033604}. Finite-temperature effects have been included by using semiclassical approximations to find the density profiles~\cite{Amoruso1998,0953-4075-35-4-102}. There are proposals of using atomic Bose-Fermi mixtures to simulate supersymmetry~\cite{Snoek05,Yu08} and quantum chromodynamics~\cite{Maeda09} systems. One interesting property of the binary atomic Bose-Fermi mixture is that at low temperatures, the bosons form BEC in a broken-symmetry phase while the fermions are in a normal, symmetric phase. The BEC transition adds additional features and challenges to the study of stable structures of Bose-Fermi mixtures.

Here we implement a theoretical framework capable of describing both fermions and bosons in a broad range of temperature and interactions to investigate the properties and structural transition of binary atomic Bose-Fermi mixtures. The framework is based on the path-integral method with the large-N expansion from quantum field theory, which has been applied to bosons~\cite{PhysRevLett.105.240402,ChienPRA12} and fermions~\cite{PhysRevA.83.053622}, respectively. For interacting bosons, the theoretical results compared favorably with the experimental data of trapped Bose gases~\cite{PhysRevA.93.033637}. Moreover, the bosonic theory predicts a second-order BEC transition, distinguishing the theory from others predicting an artificial first-order transitions~\cite{PhysRevLett.105.240402}. For two-component fermions with attractive interactions, the framework shows the BCS-Leggett theory of the BCS-BEC crossover can be viewed as a leading-order large-N expansion~\cite{PhysRevA.83.053622}. One important feature of the large-N theory is its well-defined free energy, the grand potential in thermodynamics~\cite{PhysRevLett.105.240402,schroeder2013introduction}, which will be used to determine the thermodynamics and structures of binary Bose-Fermi mixtures in box or harmonic potentials.

We remark that there have also been intense studies on atomic Bose-Fermi superfluid-superfluid mixtures~\cite{0953-4075-50-1-01LT01,Ferrier-Barbut1035,PhysRevLett.117.145301,PhysRevLett.118.055301}. However, superfluids of fermions require two components to form the Cooper pairs, and adding the bosons then requires at least three components of atoms in the mixture. Here we focus on binary boson-fermion mixtures, but the theoretical framework can be generalized to superfluid mixtures as well. Since thermodynamics and structures are the main focuses in the study of atomic mixtures, there are also relevant studies of boson-boson mixtures~\cite{PhysRevLett.81.1539,ChienPRA12} and fermion-fermion mixtures~\cite{PhysRevLett.110.165302,PhysRevA.90.023605,1367-2630-18-1-013030,Jo1521}. However, the different spin-statistics make binary Bose-Fermi mixtures particularly interesting and challenging.

The paper is organized as follows. Section~\ref{sec:theory} outlines the theoretical path-integral framework of binary atomic Bose-Fermi mixtures. It shows how to construct the thermodynamic free energy and obtain the thermodynamic quantities from the free energy. Section~\ref{sec:PS} shows how to identify phase separation from the self-intersection of the free energy curve. 
Section~\ref{sec:box} presents the construction of phase separation of the atomic mixtures in a box potential by using the lever rule. The phase diagrams for selected parameters at zero and finite temperatures are presented. Section~\ref{sec:harmonic} shows the local density approximation of atomic Bose-Fermi mixtures in harmonic traps. Typical density profiles of the mixtures are also presented. In this work we show the results of $^6$Li-$^7$Li and $^6$Li-$^{41}$K mixtures, but the framework should be applicable to other binary Bose-Fermi mixtures as well. Finally, Section~\ref{sec:conclusion} concludes our work. The Appendix summarizes the lever rule for constructing phase separation structures in equilibrium.

\section{Path integral formalism of boson-fermion mixtures}\label{sec:theory}
The Hamiltonian of a binary Bose-Fermi mixture is
\begin{eqnarray}
H&&=\int d^3x \left(-\frac{1}{2}\left(\phi^*\frac{\hbar^2\nabla^2}{2m_B}\phi+\phi\frac{\hbar^2\nabla^2}{2m_B}\phi^*\right) \right.\nonumber\\
&&-\frac{1}{2}\left(\psi^*\frac{\hbar^2\nabla^2}{2m_F}\psi+\psi\frac{\hbar^2\nabla^2}{2m_F}\psi^*\right)\nonumber\\
&&\left.+\frac{1}{2}\lambda_{BB}\left(\phi^*\phi\right)^2+\lambda_{BF}\psi^*\psi\phi^*\phi\right).
\end{eqnarray}
Here $\phi$ and $\psi$ are the bosonic and fermionic fields, $m_B$, $m_F$, $\lambda_{BB}$, and $\lambda_{BF}$ are the boson mass, fermion mass, boson-boson coupling constant, and boson-fermion coupling constant, respectively. The coupling constants can be written as $\lambda_{BB}=4\pi \hbar^2 a_{BB}/m_B$, and $\lambda_{BF}=2\pi \hbar^2 a_{BF}(m_B+m_F)/m_B m_F$, where $a_{BB}$ ($a_{BF}$) is the boson-boson (boson-fermion) two-body $s$-wave scattering length. Since there is only one component of fermions, there is no fermion-fermion interactions because the Pauli exclusion principle suppresses the $s$-wave scattering between identical fermions. In what follows, we set $\hbar=k_B=1$. 

Using the imaginary time formalism with $\tau=-it$, the corresponding Euclidean Lagrangian density can be found. After including the chemical potentials $\mu_B$ and $\mu_F$ for the bosons and fermions and the source term for the bosonic fields, we obtain
\begin{eqnarray}
\mathcal{L}_E=&&\frac{1}{2} \Phi ^{\dagger} G_{0 B}^{-1} \Phi -\mu _B \phi ^* \phi + \frac{1}{2} \lambda _{{BB}} \left(\phi ^* \phi \right)^2-J^{\dagger} \Phi\nonumber\\
 +&&\frac{1}{2} \Psi ^{\dagger} G_{0 F}^{-1} \Psi -\mu _F \psi ^* \psi +\lambda _{{BF}} \psi ^* \psi  \phi ^* \phi .
 \label{eq:Lagrangian}
\end{eqnarray}
where $G_{0B}^{-1}=diag\left( \partial _{\tau }+\frac{\nabla ^2}{2 m_B}, - \partial _{\tau }+\frac{\nabla ^2}{2 m_B}\right)$, $G_{0F}^{-1}=diag\left( \partial _{\tau }+\frac{\nabla ^2}{2 m_F}, - \partial _{\tau }+\frac{\nabla ^2}{2 m_F}\right)$, $\Phi =\left(
\begin{array}{cc}
\phi,  & \phi ^* \\
\end{array}
\right)^{\mathsf{T}}$, $\Psi =\left(
\begin{array}{cc}
\psi,  & \psi ^* \\
\end{array}
\right)^{\mathsf{T}}$, and $J=\left(
\begin{array}{cc}
j, & j^* \\
\end{array}
\right)^{\mathsf{T}}$. Here $J$ is the source coupled to $\Phi$ and the superscript $T$ denote the transpose of a matrix.
The grand partition function is then
\begin{eqnarray}
Z=\int \mathcal{D}\Phi \mathcal{D}\Psi e^{-S_E},
\end{eqnarray}
where $S_E=\int d^4x_E\mathcal{L}_E$ is the Euclidean action, and $d^4x_E=d\tau d^3x$. From now on we will drop the subscript $E$.

Since the fermion contributions are quadratic, they can be integrated out similar to the Peierls transition problem~\cite{zee2010quantum,Schakel_book}. Afterwards, the action becomes an effective action for the bosons:
\begin{eqnarray}
S_{B}[J,\Phi]&&=\int d^4x \left(\frac{1}{2} \Phi ^{\dagger} G_{0 B}^{-1} \Phi -\mu _B \phi ^* \phi + \frac{1}{2} \lambda _{{BB}} \left(\phi ^* \phi \right)^2\right. \nonumber\\
-J^{\dagger} \Phi
&&\left.-\frac{1}{2} \operatorname {tr} \ln G_F^{-1}\right),
\end{eqnarray}
where $G_{F}^{-1}=G_{F0}^{-1}+[-\mu_F+\lambda _{BF}\phi^*\phi]\bar{1}$ and $\bar{1}$ is the $2\times 2$ identity matrix.

To handle the boson-boson interactions, we implement the large-N expansion similar to Refs.~\cite{PhysRevLett.105.240402,ChienPRA12,PhysRevA.93.033637}. 
The idea behind the large-N expansion for bosons is to introduce $N$ fictitious copies of the original field and assume an internal SU(N) symmetry. By scaling the interaction strength properly and introducing a composite field representing $\phi ^* \phi$, the effective action and the partition function can be expanded according to powers of $1/N$. An approximation is then obtained by truncating the theory at the leading order and set $N=1$ in the final expression. There are at least two ways of introducing the composite field. One can either use the Hubbard-Stratonovic transformation when the interaction is quartic~\cite{PhysRevLett.105.240402} or use the Dirac delta functional~\cite{ChienPRA12}. For quartic interaction terms the two methods are equivalent. Here we follow the latter and introduce
$1
=\frac{1}{\mathcal{N}}\int \mathcal{D}\alpha \mathcal{D}\chi~ e^{\int d^4x\frac{\chi  }{\lambda _{{BB}}}\left(\alpha -\lambda _{{BB}} \phi ^* \phi \right)}$ to the partition function.
Here $\mathcal{N}$ is a normalization constant and the $\chi$ integration runs parallel to the imaginary axis \cite{ChienPRA12}. Then, the action can be decomposed into a collection of quadratic terms of the bosonic field: 
\begin{eqnarray}
S_{B}[\Phi,J,\Psi,\chi,\alpha]
&&
=\int d^4x \left(\frac{1}{2} \Phi ^{\dagger} G_B^{-1} \Phi -\frac{\mu _B \alpha }{\lambda _{{BB}}}+ \frac{\alpha ^2}{2 \lambda _{{BB}}}\right. \nonumber\\
&&\left. -J^{\dagger} \Phi-\frac{\chi  \alpha }{\lambda _{{BB}}}
-\operatorname {tr} \ln G_F^{-1}\right),
\end{eqnarray}
where $G_{B}^{-1}=G_{B0}^{-1}+\chi\bar{1}$ and $G_{F}^{-1}=G_{F0}^{-1}+[-\mu_F+(\lambda _{BF}/\lambda_{BB})\alpha]\bar{1}$. It is customary to introduce the sources coupled to $\chi$ and $\alpha$~\cite{ChienPRA12}, so we include the terms $-(s\alpha+g\chi)$.

Now the action is quadratic in the bosonic field $\phi$, so we integrate out $\phi$ and obtain $Z=\int\mathcal{D}\chi\mathcal{D}\alpha \exp[-S_{eff}]$. The effective action is
\begin{eqnarray}
S_{eff}[J,s,g]&&=\int d^4x \left(\frac{1}{2} \operatorname{tr}G_B^{-1}-\frac{\mu _B \alpha }{\lambda _{{BB}}}+ \frac{\alpha ^2}{2 \lambda _{{BB}}}-\frac{\chi  \alpha }{\lambda _{{BB}}} \right.\nonumber\\
&&\left. -\frac{1}{2} J^{\dagger} G_B J
-\frac{1}{2} \operatorname{tr}\ln G_F^{-1}-s\alpha-g\chi \right).
\end{eqnarray}
However, $S_{eff}$ is a functional of the sources $j, s$, and $g$, therefore it is inconvenient for deriving thermodynamic relations. It has been shown that~\cite{PhysRevLett.105.240402,ChienPRA12} a Legendre transform of $S_{eff}$ gives the grand potential (the subscript $c$ denotes the expectation value)
\begin{eqnarray}
\Gamma=\int d^4x (J^\dagger\Phi_c+g\chi_c+s\alpha_c)+S_{eff}.
\end{eqnarray}
Importantly, $\Gamma$ is a functional of the expectations of $\phi$, $\alpha$, and $\chi$, therefore it is suitable for studying thermodynamics. Moreover, we have the relation $\delta\Gamma/\delta \Phi_c^*=\int G_B^{-1}\Phi_c=J$. For static homogeneous fields, we define the effective potential $V_{eff}=\Gamma/(V\beta)$, where $V$ is the volume and $\beta=(k_B T)^{-1}$. In equilibrium, the effective potential is the volume density of the grand potential~\cite{schroeder2013introduction} in thermodynamics, and it is related to the pressure by $V_{eff}=-p$~\cite{ChienPRA12}. In the following we will drop the subscript $c$ and focus on the expectation values. 

To the leading order of $1/N$, the effective potential is
\begin{eqnarray}
V_{eff}&&=\chi\phi^*\phi-\frac{\mu _B \alpha }{\lambda _{{BB}}}+ \frac{\alpha ^2}{2 \lambda _{{BB}}}-\frac{\chi  \alpha }{\lambda _{{BB}}}\nonumber\\
&&+\frac{1}{2} \operatorname{tr}G_B^{-1}-\frac{1}{2} \operatorname{tr}G_F^{-1}.
\end{eqnarray}
Here we have set $N=1$ to match the atomic gas.
After summing up the Matsubara frequencies~\cite{nair2006quantum}, the last two terms become
$\frac{1}{2} \operatorname{tr}\ln G_B^{-1}
=\sum _q \left(\frac{\omega _{B }}{2}+\frac{1}{\beta }\ln \left(1-e^{-\beta  \omega _{B}}\right)\right)$ and
$\frac{1}{2} \operatorname{tr}\ln G_F^{-1}
=\sum _k \left(\frac{\omega _{F}}{2}+\frac{1}{\beta }\ln \left(1+e^{-\beta  \omega _{F}}\right)\right)$,
where $\omega _{B}=\frac{q^2}{2 m_B}+\chi$ and $\omega _{F}=\frac{k^2}{2 m_F}-\mu _F+\frac{\lambda _{{BF}}}{\lambda _{{BB}}}\alpha$. Since the contact potential introduces infinities in the integrals, we follow the standard renormalization procedure~\cite{PhysRevLett.105.240402,ChienPRA12} and obtain the renormalized effective potential 
\begin{eqnarray}\label{eq:VeffRenorm}
V_{eff}&&=\chi\phi^*\phi-\frac{\mu _B \alpha }{\lambda _{{BB}}}+ \frac{\alpha ^2}{2 \lambda _{{BB}}}-\frac{\chi  \alpha }{\lambda _{{BB}}}\nonumber\\
&&+\sum _q \frac{1}{\beta }\ln \left(1-e^{-\beta  \omega _{B}}\right)-\sum _k \frac{1}{\beta }\ln \left(1+e^{-\beta  \omega _{F}}\right).
\end{eqnarray}

The equations of state are obtained by minimizing the effective potential
\begin{eqnarray}
\frac{\partial V_{eff}}{\partial \phi^*}=\chi  \phi =0.
\end{eqnarray}
\begin{eqnarray}
\frac{\partial V_{eff}}{\partial \alpha}
&&=-\frac{\mu _B}{\lambda_{BB}}+ \frac{\alpha}{\lambda_{BB}}-\frac{\chi}{\lambda_{BB}}+\frac{\lambda _{{BF}}}{\lambda_{BB}} \sum_k n_B	(\omega_{F})\nonumber\\
&&=0.
\label{eq:LargeN_chi}
\end{eqnarray}
\begin{eqnarray}
\frac{\partial V_{eff}}{\partial \chi}
=\phi ^* \phi-\frac{\alpha }{\lambda _{{BB}}} +\sum _q n_B (\omega _{B})
=0.
\label{eq:LargeN_rho_B}
\end{eqnarray}
There are additional relations from thermodynamics:
\begin{eqnarray}
-\frac{\partial V_{eff}}{\partial \mu_B}=\frac{\alpha }{\lambda _{{BB}}}=\rho _B.
\end{eqnarray}
\begin{eqnarray}
-\frac{\partial V_{eff}}{\partial \mu_F}=\sum _k n_F (\omega _{F})=\rho _F.
\end{eqnarray}
Here $n_B$ ($n_F$) is the Bose (Fermi) distribution function, and $\rho_B$ ($\rho_F$) is the boson (fermion) density. 
By solving the equations of state and finding the corresponding $V_{eff}$ when the parameters are varied, we will show how to map out the stability of binary atomic Bose-Fermi mixtures in the next section.

We remark that here we only introduce an auxiliary field representing the normal density in the large-N expansion. By introducing two auxiliary fields representing the normal and anomalous densities, the theory is called the leading-order auxiliary field (LOAF) theory~\cite{PhysRevLett.105.240402}. The LOAF theory naturally recovers the Bogoliubov theory at low temperatures when the interaction is weak, but it is not fully compatible with the local density approximation in the strongly interacting regime when dealing with harmonically trapped Bose gases~\cite{PhysRevA.93.033637}.

\section{Phase separation and structural transition}\label{sec:PS}
By examining the kinetic and interaction energies of the bosons and fermions, it has been argued \cite{PhysRevA.61.053605} there is a structural transition. Across the transition, a miscible state will transform into phase separation, where two phases with different ratios of fermions and bosons coexist. The phase separation has been observed in recent experiments~\cite{Lous18}.
After obtaining the effective potential, we elucidate the thermodynamics behind the structural transition. Firstly, we remark that Eq.~\eqref{eq:VeffRenorm} is the free energy of a miscible mixture. However, it will show instabilities where phase separation should be constructed in the parameter space. 

\begin{figure}[th]
	\begin{center}
		\includegraphics[width=1\linewidth]{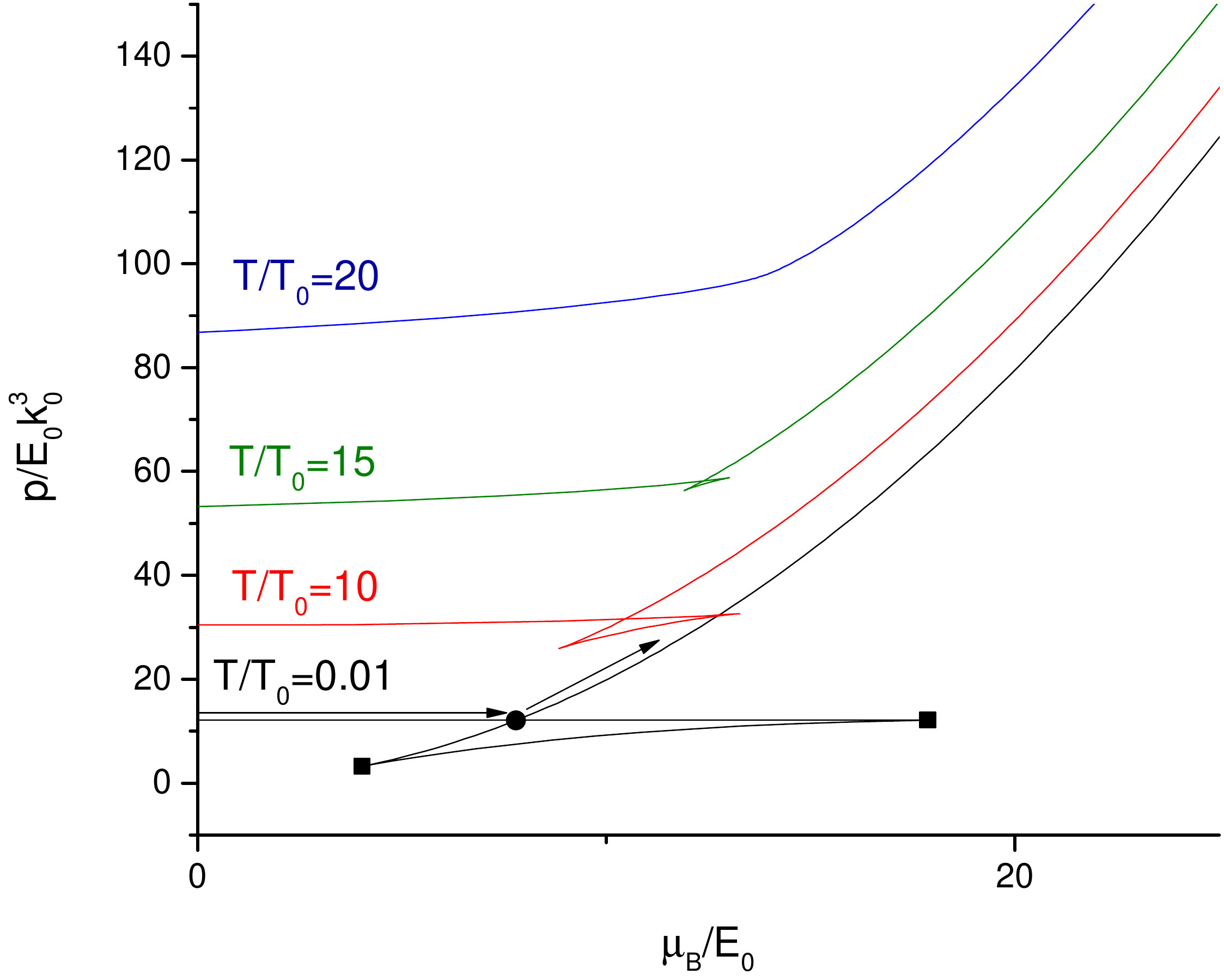}
		\caption{The pressure $p$ of a uniform binary Bose-Fermi mixture as a function of the boson chemical potential $\mu_B$. We fix $a_{BB} k_0=0.1$, $a_{BF} k_0=0.5$, and $\mu_F/E_0=20$. The corresponding temperature is labeled next to each curve. Here $k_0^3=(N_F+N_B)/V$, $E_0=\hbar^2 k_0^2/(2m_F)$, and $k_B T_0=E_0$. At high temperatures (for example, $T/T_0=20$), $p$ increases monotonically with $\mu_B$. At low temperatures, the curve exhibits a loop. The point where the curve intersects itself is where phase separation occurs, as demonstrated by the black dot on the $T/T_0=0.01$ curve. After following the Maxwell construction, the system moves from one phase to the phase separation point and then to another phase, as indicated by the arrows shown on the $T/T_0=0.01$ curve without traversing the loop. The two vertices in the loop (marked by the black squares on the $T/T_0=0.01$ curve) correspond to the spinodal points.}
		\label{fig:1field_muB_vs_p}
	\end{center}
\end{figure}

Figure~\ref{fig:1field_muB_vs_p} shows $p=-V_{eff}$ as a function of $\mu_B$ for different values of $T$ with fixed $a_{BB}, a_{BF}, \mu_F$. Here the solution of the equations of state has been found numerically. The units $k_0$ and $E_0$ are fixed by external scales from the trapping potentials. We will specify those units in the discussions of the box and harmonic potentials. At high temperatures the $p$ curve is smooth and monotonic. However, at low temperatures the curve can form a loop and intersect itself. Such a loop structure in the free energy is a typical example of a first-order transition in thermodynamics~\cite{schroeder2013introduction,kittel1980thermal,reichl2009modern}.

Although the free energy (in this case the effective potential) exhibits a loop, the equilibrium system does not traverse the loop. Instead, following the Maxwell construction~\cite{schroeder2013introduction,kittel1980thermal,reichl2009modern} the system transforms from one phase to another by going through a phase coexistence (phase separation) point indicated by where the free energy curve intersects itself. The ratios between the bosons and fermions in the two phases at the phase separation point can also be inferred by the two intersecting lines at the intersection. Therefore, by analyzing the free energy curve and performing the Maxwell construction if a loop is found, the stable structure at each point in the parameter space can be mapped out in a systematic way. We mention that although the interior of the free-energy loop cannot be traversed by a real system in equilibrium, it offers additional information. For instance, the vertices of the loop are the spinodal points separating different types of dynamics when the system is driven into the phase separation regime~\cite{chaikin1995principles}.

One also observes that at the intersecting point of the free-energy curve, the two coexisting phases have the same pressure and chemical potentials of the two species. Those conditions guarantee mechanical and chemical equilibrium~\cite{tester1997thermodynamics,PhysRevA.61.053605}. However, the densities usually differ in the two phases, which is a common feature of phase separation.
If the total particle numbers of the bosons and fermions in the initial unstable mixture are known instead of the chemical potentials, one can construct the stable structures by finding the correct volume ratio according to the level rule, which balances the extensive variables. (See the Appendix for a summary of the lever rule.)  In the following sections, we will show how to map out the structures and phase diagrams by examining the behavior of $V_{eff}$. We will discuss two types of confining potentials, the box~\cite{Gaunt2013,PhysRevLett.112.040403,PhysRevLett.119.190404} and harmonic~\cite{pethick2008bose,pitaevskii2003bose} potentials, commonly used in cold-atom experiments.

\begin{figure}[th]
	\includegraphics[width=1\linewidth]{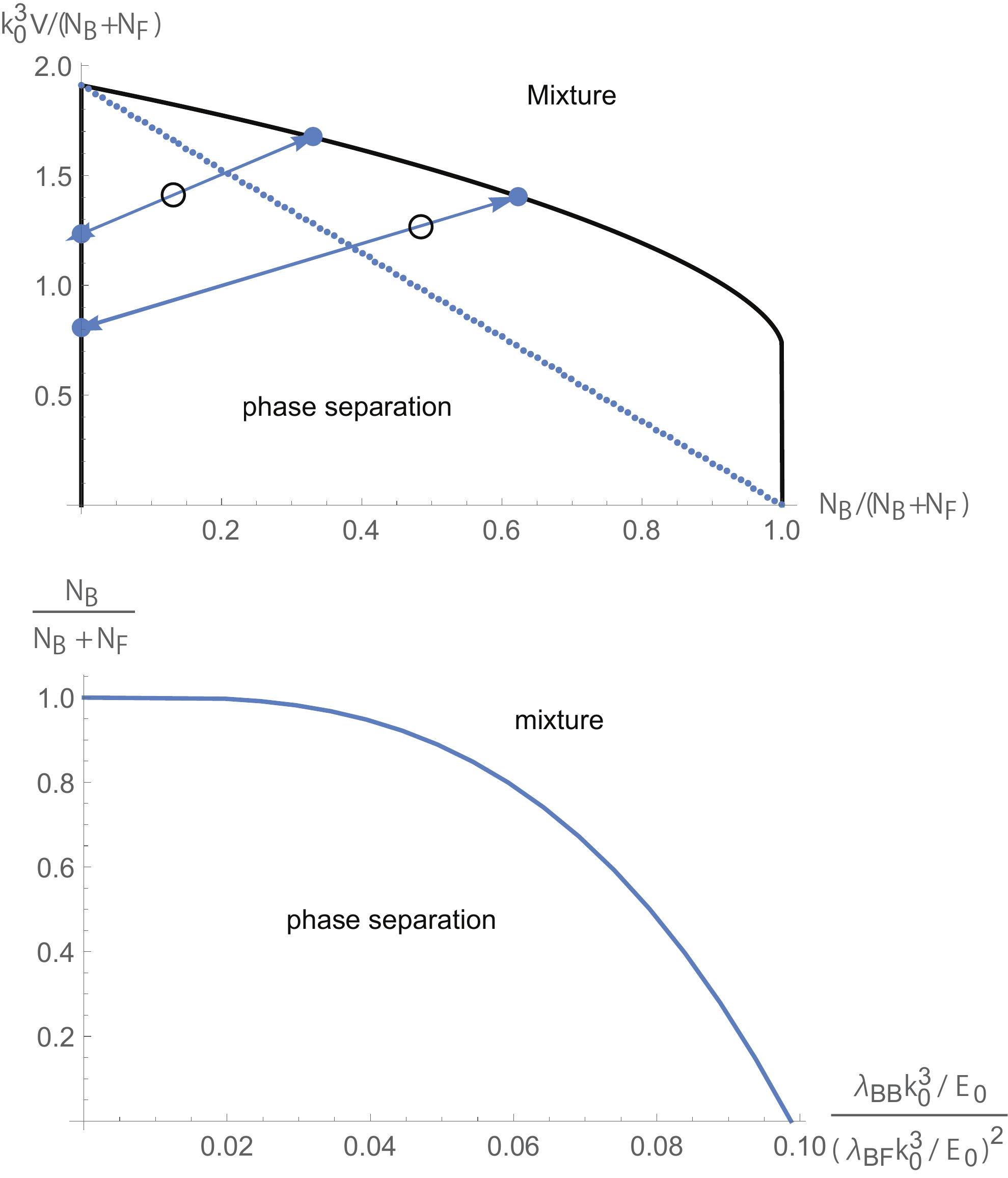}
	\caption{(Top panel) Zero-temperature phase diagram of an equal-mass binary Bose-Fermi mixture. Here $1/k_0$ is fixed by an external length scale, $\lambda_{BB} k_0^3/E_0/\left(\lambda_{BF} k_0^3/E_0\right)^2=1/4\pi$, and $k_0 a_{BF}=0.5$. $N_B$ ($N_F$) is the total number of bosons (fermions). The solid lines are the phase boundary of phase separation. Above the boundary the system is a stable mixture and below it phase separation occurs. The dotted line is the spinodal line. Two examples of phase separation are shown by the circles and arrows. The circles show the initial (unstable) mixture compositions, and the systems will reach equilibrium by separating into the solid dots that are collinear with the circles. The concentrations of the separated phases can be found from the solid dots. (Bottom panel) Fraction of bosons in the boson-rich phase of a system in the phase separation regime at zero temperature with $k_0^3 V/(N_B+N_F)=1$. As shown in the top panel, the other phase is a pure-fermion phase. The fraction of bosons in the boson-rich phase decreases as the boson-boson interaction increases. For the parameters selected here, the mixture is stable against phase separation if $(\lambda_{BB} k_0^3/E_0)/\left(\lambda_{BF} k_0^3/E_0\right)^2 > 0.099$.}
	\label{fig:zero_temperature}
\end{figure}

\section{Box potential}\label{sec:box}
The realization of optical box potentials for trapping cold-atoms~\cite{Gaunt2013,PhysRevLett.112.040403,PhysRevLett.119.190404} allows a direct comparison between theories derived for uniform gases and experiments.
For a binary Bose-Fermi mixture in a box potential, the mixture will separate into a boson-rich phase and a fermion-rich phase if the boson-fermion interaction is strong and the temperature is low. In such a system, the fixed variables are the volume $V$ and particle numbers $N_B$ and $N_F$. Here we choose $m_0$ as the mass of ${}^6$Li.  The box volume $V_0$ in Ref.~\cite{Gaunt2013} is chosen as the unit of volume, and the relation $V_0=k_0^{-3}$ gives the length unit $1/k_0\approx 44.1 nm$. The energy and temperature units for the box potential are $E_0=\hbar^2 k_0^2/(2m_0)$ and $T_0=E_0/k_B$, respectively.
Given the temperature $T$ and interaction strengths, the stable equilibrium corresponds to the minimum of the Helmholtz free energy
$\mathcal{A}=V_{eff}V+\sum_{i=B,F} \mu_i N_i$. Importantly, the phase separation point is independent of which free energy one uses to locate it as long as the Legendre transform is implemented correctly.

To find out the volume ratio of the atomic mixture in phase separation, we use the lever rule~\cite{kittel1980thermal,schroeder2013introduction} that determines the balance of the extensive variables. An important difference between gaseous mixtures and liquid or solid mixtures drastically complicates how to apply the lever rule for phase separation. In liquid or solid mixtures, the density of each constituent is constant. For example, the density of water and the density of phenol in a phase separation structure are indistinguishable from their densities in a miscible mixture~\cite{doi:10.1021/ja01291a001}. Assuming there are two species $1$ and $2$, if the total particle fraction $x=N_1/(N_1+N_2)$ is specified and the two species are incompressible, the lever rule of the Helmholtz free energy is identical to that of the Gibbs free energy for liquid or solid mixtures~\cite{kittel1980thermal}. Hence, the fraction in each separated phase can be determined straightforwardly. However, atomic gases are compressible and their densities can be flexible. As a consequence, it is not enough to search for the minimum of $\mathcal{A}$ by varying $x$ only. Instead, one has to apply the lever rule to the Helmholtz free energy by considering possible changes of $x$ as well as $V_{\alpha}/V_{\beta}$, where $V_{\alpha}$ and $V_{\beta}$ denote the volumes of the two separated phases. The minimization problem in a two-dimensional parameter space is much more complicated than the conventional minimization of liquid or solid mixtures.

Nevertheless, the complication can be circumvented by using the effective potential and locating the loop to map out the phases intersected at the phase-coexistence point, as illustrated in Fig.~\ref{fig:1field_muB_vs_p}. To match the fixed volume and total particle numbers, we construct the phase boundary by tuning the extensive variables so that their sums from the separated phases match the original mixture according to the lever rule (see the Appendix for details). Specifically, the initial condition of the system has fixed numbers of bosons $N_{Bo}$ and fermions $N_{Fo}$, and the total volume $V_{o}$ is also fixed. Here the subscript $o$ denotes the quantities in the initial (unstable) mixture. The conservation of extensive variables impose the following constraints, known as the lever rules. 
\begin{eqnarray}
\sum_i \rho_{Bi} v_i=\rho_{Bo},~
\sum_i \rho_{Fi} v_i=\rho_{Fo},~
\sum_i v_{i} =1.
\label{eq:extensive_variable_conservation_Helmholtz}
\end{eqnarray}
Here the subscript $i$ denotes the quantities in the $i$th phase when the system is in phase separation, and $v_i=V_i/V_o$ is the volume fraction of the $i$th phase.

\begin{figure}[ht!]
	\includegraphics[width=1\linewidth]{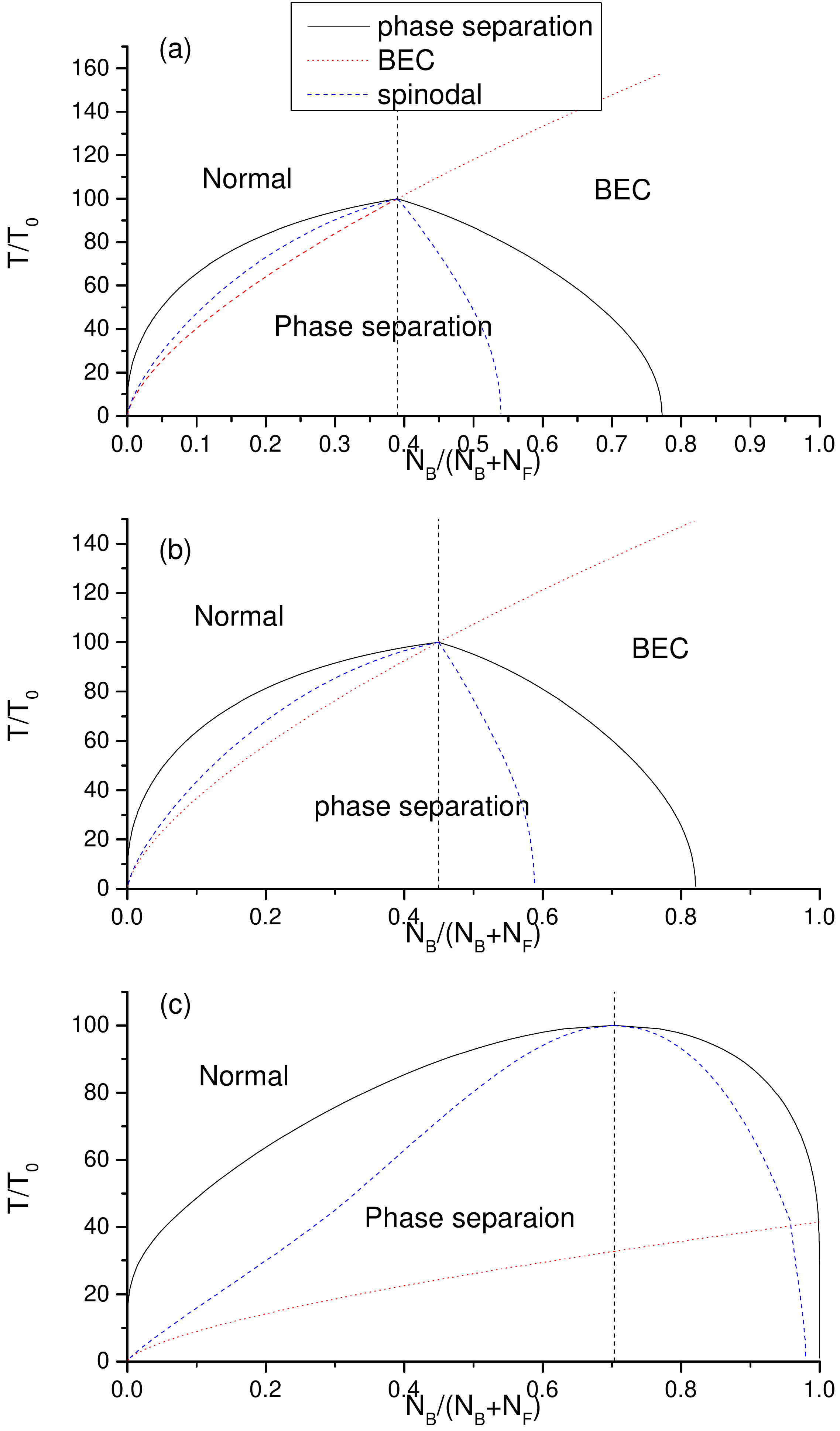}
	\caption{(a) Phase diagram of an equal-mass Bose-Fermi mixture in a box potential with a fixed volume and total boson fraction $N_B/\left(N_B+N_F\right)=0.39$ as indicated by the vertical black dashed line. The black lines forming a dome is the phase separation boundary showing the boson fractions in the two separated phase. Here $Vk_0^3/(N_B+N_F)=0.067$, $a_{BB}k_0=a_{BF}k_0=0.1$, and $T_F/T_0=31$. For the parameters chosen here, the critical temperature of phase separation is $T_c/T_0=100$. Below $T_c$ (above $T_c$) phase separation (uniform mixture) is stable. The blue dashed curves are the spinodal lines indicating the two vertices in the loop of the free energy. The red dotted curve is the BEC transition line, under which BEC of bosons can be found. (b) Phase diagram of a $^6$Li and $^7$Li mixture with a fixed volume and $N_B/\left(N_B+N_F\right)=0.45$. Here $k_0^3 V/(N_B+N_F)=0.0061$, $k_0 a_{BB}=k_0 a_{BF}=0.1$, $T_F/T_0=31$, and $T_c/T_0=100$. (c) Phase diagram of a $^6$Li and $^{41}$K mixture with a fixed volume and  $N_B/\left(N_B+N_F\right)=0.70$. Here $k_0^3 V/(N_B+N_F)=0.0036$, $k_0 a_{BB}=k_0 a_{BF}=0.1$, $T_F/T_0=290$, and $T_c/T_0=100$.}
	\label{fig:phase_diagram_fixed_V}
\end{figure}

Figure~\ref{fig:zero_temperature}(a) shows the zero-temperature phase diagram of an equal-mass Bose-Fermi mixture with selected values of $a_{BB}$ and $a_{BF}$, respectively. The phase separation regime has a skewed dome-shape boundary. If an initially mixed state is prepared inside the phase-separation regime, it will reach equilibrium by separating into two phases with different ratios of the bosons and fermions. We illustrate this by showing two initial conditions and their corresponding phase separation compositions. It has been argued that when phase separation occurs at zero temperature, one of the phases will be of only fermions with no bosons, but the other phase does not necessarily have pure bosons only~\cite{PhysRevA.61.053605,PhysRevB.78.134517}. Our result confirms this observation. As one can see in Fig.~\ref{fig:zero_temperature}(a), the separation can be either into a boson-only phase and a fermion-only phase near the bottom of the phase separation regime, or a fermion-only phase and a partially mixed phase in the upper regime of phase separation.

Our method of using the loop of the free energy to locate and construct phase separation has the advantage that the fraction of bosons in the boson-rich phase can be evaluated as the interactions are varied. Figure~\ref{fig:zero_temperature}(b) shows the boson fraction for an equal-mass Bose-Fermi mixture in the phase separation regime at zero temperature. We recall that the other phase is a fermion-only phase. One can see that when the boson-boson interaction is weak (strong) compared to the boson-fermion interaction, the boson fraction approaches $1$ (is below $1$). When the boson-boson interaction is too strong, the mixture remains stable and no phase separation is observed. Incidentally, $^3$He-$^4$He liquid mixtures have been shown to separate into a fermion-only phase and a partially mixed phase at low temperatures~\cite{BERNTSEN1979107,EBNER197177}.

Our theoretical framework naturally applies to binary Bose-Fermi mixtures in a box potential at finite temperatures.  Fig.~\ref{fig:phase_diagram_fixed_V} (a) shows the finite-temperature phase diagram of an equal-mass binary Bose-Fermi mixture in a box potential. Fig.~\ref{fig:phase_diagram_fixed_V} (b) and (c) show the phase diagrams of $^6$Li-$^7$Li and $^6$Li-$^{41}$K mixtures, respectively. The phase diagrams are specific to a selected total boson fraction because we construct the phase-separation boundary by locating the self-intersecting point of the effective potential and then tuning the extensive variables to match the selected total boson fraction according to the lever rule. We select the boson ratios so that the different mixtures with different mass ratios shown in Fig.~\ref{fig:phase_diagram_fixed_V} all start to phase separate around $T/T_0=100$. If a different boson ratio is used, the dome-shape boundary will only change quantitatively. 

For atomic mixtures, the individual phases in the phase separation structure do not necessarily have the same densities as the initial unstable mixture because the gaseous phases adjust their densities to reach the lowest total free energy. Thus, the full phase diagram would require a 3D plot. However, if the pressure is fixed, or if the change in the densities are negligible like conventional liquid or solid mixtures, a 2D plot would be sufficient. Here, we do not fix the pressure because it is unphysical to fix both the pressure and volume of a compressible gas. Therefore, Fig.~\ref{fig:phase_diagram_fixed_V} does not explicitly specify the densities of the separated phases. If one needs the density of each phase, a full 3D plot with the overall density as the third axis can be constructed. Nevertheless, Fig.~\ref{fig:phase_diagram_fixed_V} is a 2D projection of the full plot showing the correct boson fraction in each phase. Importantly, the construction guarantees the system in both mechanical and diffusive equilibrium.

The BEC transition temperature is also presented for each case in Fig.~\ref{fig:phase_diagram_fixed_V}. The leading-order large-N theory of a uniform Bose gas leads to the same BEC transition temperature as the noninteracting Bose gas~\cite{ChienPRA12}. Here the BEC transition line is obtain by using the boson density in the corresponding phase if the system is in the phase separation regime. Below (above) the transition temperature, the phase has (has no) BEC of the bosons. For a homogeneous Bose gas,
\begin{eqnarray}
	\displaystyle \frac{T_{c}}{T_0}=\left({\frac {N_B \rho /k_0^3 }{\zeta (3/2) \left(N_B+N_F\right)}}\right)^{2/3}{\frac {4\pi }{m_B/m_0}},
\end{eqnarray}
where $\zeta(y)$ is the Riemann zeta function. The Fermi temperature of single-component, homogeneous and noninteracting fermions is
\begin{eqnarray}
\frac{T_F}{T_0} = \frac{1}{m_F/m_0} \left( 6 \pi^2 \rho_{F}/k_0^3 \right)^{2/3} .
\end{eqnarray}
The Fermi temperature of Fig.~\ref{fig:phase_diagram_fixed_V} is given by a noninteracting Fermi gas with the same fermion mass and density.

As one can see in Fig.~\ref{fig:phase_diagram_fixed_V}, only one branch of the phase separation boundary can have BEC of the bosons. We found this to be a generic feature from our theory. When the mass difference between the fermions and bosons are large, such as the case shown in Fig.~\ref{fig:phase_diagram_fixed_V}(c), BEC can only be found at relatively low temperatures on one of the phase separation boundary.
We also show the spinodal lines inside the phase separation regime by locating the vertices of the loop of the free energy. The spinodal lines imply divergence of the density susceptibilities $\left(\partial \rho_i/\partial \mu_j\right)_{T,\mu_{k \ne j}}$, where $i,j,k=B,F$, if one assumes the system remains miscible. In equilibrium, however, the spinodal lines are preempted by phase separation.

\section{Harmonic potential}\label{sec:harmonic}
For binary atomic Bose-Fermi mixtures in harmonic traps, we use the local density approximation (LDA)~\cite{pethick2008bose,RevModPhys.71.463} to approximate the inhomogeneous density profiles. The assumption behind the LDA is that the trap potential varies slowly so that each point in the trap can be treated as a homogeneous system with an effective local chemical potential for each species. After obtaining the physical quantities at different points, the overall physical quantities can be found by an integration over the trap. The leading-order large-N theory of interacting bosons has been shown to be compatible with the LDA~\cite{PhysRevA.93.033637}. Here, we assume the harmonic traps for the bosons and fermions are isotropic with trap frequencies $\omega_{oB}$ and $\omega_{oF}$, respectively. The trap centers are assumed to be at the same location. For harmonically trapped Bose-Fermi mixtures, we still choose the mass unit $m_0$ as the mass of $^{6}$Li. The length unit is given by the fermion harmonic length, $a_{HF}=\sqrt{\hbar/m_F\omega_{oF}}\equiv 1/k_0$. We take a typical value of $\omega_{oF}=2\pi\times140Hz$ from Ref.~\cite{PhysRevA.95.053627}, and it translates to $1/k_0=3.45 \mu m$. Moreover, we choose the boson harmonic length and fermion harmonic length equal to each other, $a_{HB}=a_{HF}$, where $a_{HB}=\sqrt{\hbar/m_B\omega_{oB}}$. The energy unit is given by $E_0=\frac{1}{2}\hbar \omega_{oF}$, which also determines the temperature unit $T_0=E_0/k_B$.

\begin{figure}
	\includegraphics[width=0.9\linewidth]{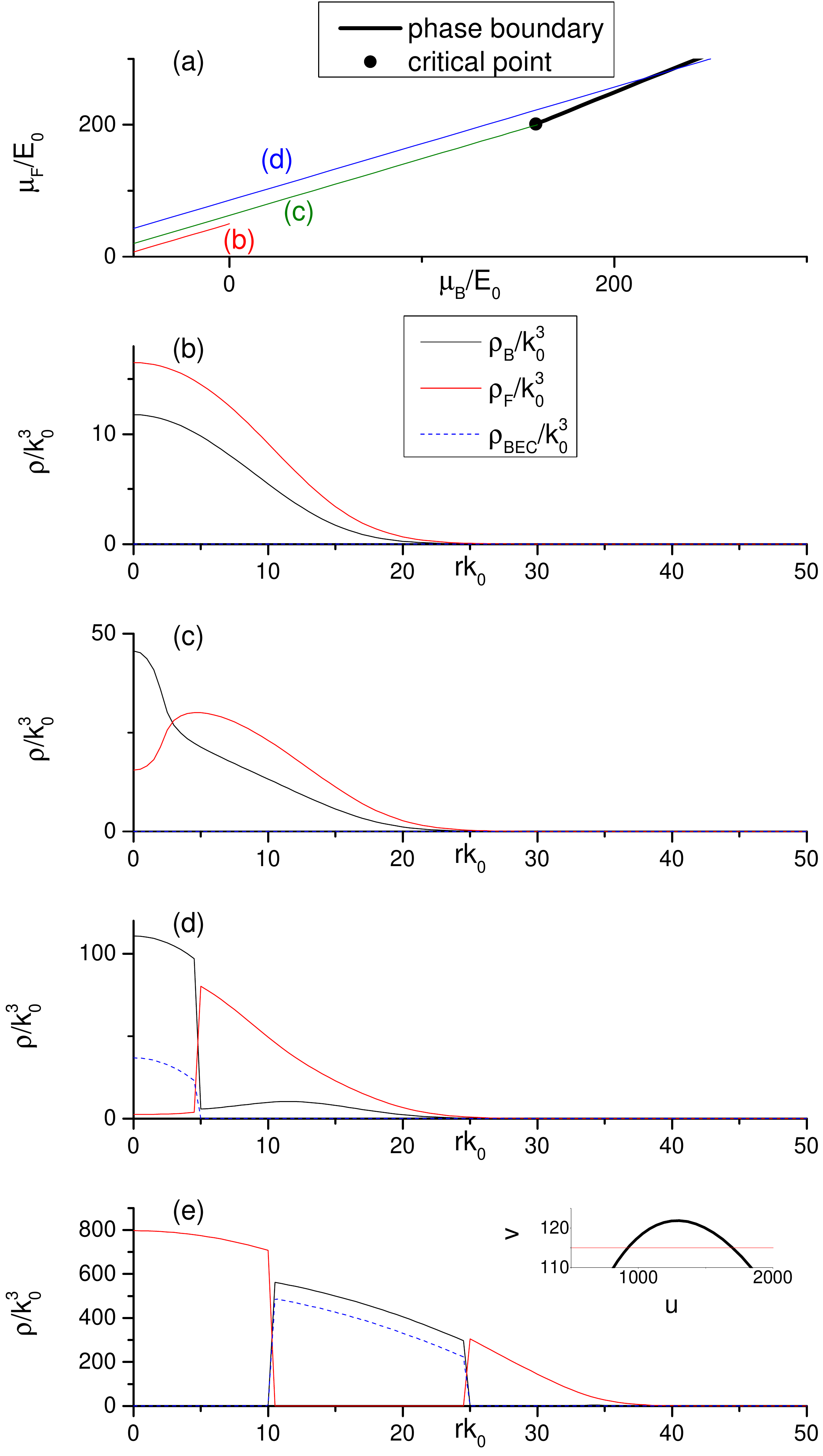}
	\caption{Harmonically trapped atomic $^6$Li - $^7$Li mixtures. (a) Phase separation boundary (thick line) and selected cases of the effective local chemical potentials (thin lines) according to the LDA. On the left (right) of the phase boundary is a fermion-rich (boson-rich) phase. The density profiles of the three lines labeled by (b), (c), and (d) are shown below. Here $T/T_0=100$, $a_{BB}k_0=0.1$, and $a_{BF}k_0=0.2$ for all panels. The solid black, solid red, and dashed blue lines are the boson density, fermion density, and boson condensate density, respectively. (b) The number of bosons (fermions) is $N_B=1.15\times 10^5$ ($N_F=2.19\times10^5$). The densities of bosons and fermions both decrease monotonically from the trap center. (c) $N_B=2.31\times 10^5$ and $N_F=4.33\times 10^5$. The fermions are pushed away from the trap center, but the local chemical potentials have not crossed the phase boundary, and the density profiles are smooth. (d) $N_B=2.95\times 10^5$ and $N_F=9.28\times 10^5$. The local chemical potentials cross the phase boundary once, and the boson-rich region discontinuously changes to the fermion-rich region. (e) Here, $N_B=2.45\times 10^7$ and $N_F=2.02\times 10^7$. The density profiles have a boson-rich region sandwiched between two fermion-rich regions because the local chemical potentials cross the phase boundary twice. The inset shows the phase boundary (thick line) and the effective chemical potentials (thin line) of (e). Here, $u=0.65 \mu_B/E_0-0.76 \mu_F/E_0$ and $v=0.76\mu_B/E_0+0.65\mu_F/E_0$.
	}
	\label{fig:phase_boundary_and_density_profile_Li6_Li7}
\end{figure}

The local chemical potentials in the LDA are given by $\mu_B(r)=\mu_{B}(r=0)-\frac{1}{2}m_B \omega_{oB}^2 r^2 $ and $\mu_F(r)=\mu_{F}(r=0)-\frac{1}{2}m_F \omega_{oF}^2 r^2 $. Using them to solve the equations of state at given $T$, $a_{BB}$, and $a_{BF}$, we obtain the densities $\rho_{B,F}(r)$ at radius $r$ in the traps. The boson condensate density $\rho_{BEC}(r)$ can also be found. The total particle numbers can be obtained by using
\begin{eqnarray}
	N_{B,F}=\int dr^3 \rho_{B,F}(r).
\end{eqnarray} 
Moreover, the BEC transition temperature of a harmonically trapped noninteracting Bose gas ~\cite{pathria2011statistical} is 
$\frac{T_{c}}{T_0}=\frac{\zeta(3)^{-1/3}\hbar\omega_{oB} N_B^{1/3}/k_B}{E_0/k_B}
=
\frac{2 \left(N_B/\zeta(3)\right)^{1/3}}{\left(a_{HB} k_0\right)^2}$.
The Fermi temperature of a harmonically trapped noninteracting single-component Fermi gas~\cite{1063-7869-59-11-1129} is
$\frac{T_F}{T_0}=\frac{6^{1/3}\hbar \omega_{oF} N_F^{1/3}/k_B}{E_0/k_B}=\frac{2 \left(6N_f\right)^{1/3}}{\left(a_{HF} k_0\right)^2}$.
Here $N_B$ and $N_F$ are the total boson and fermion numbers, respectively, and $\zeta(x)$ is the Riemann zeta function.

\begin{figure}[t]
	\includegraphics[width=1\linewidth]{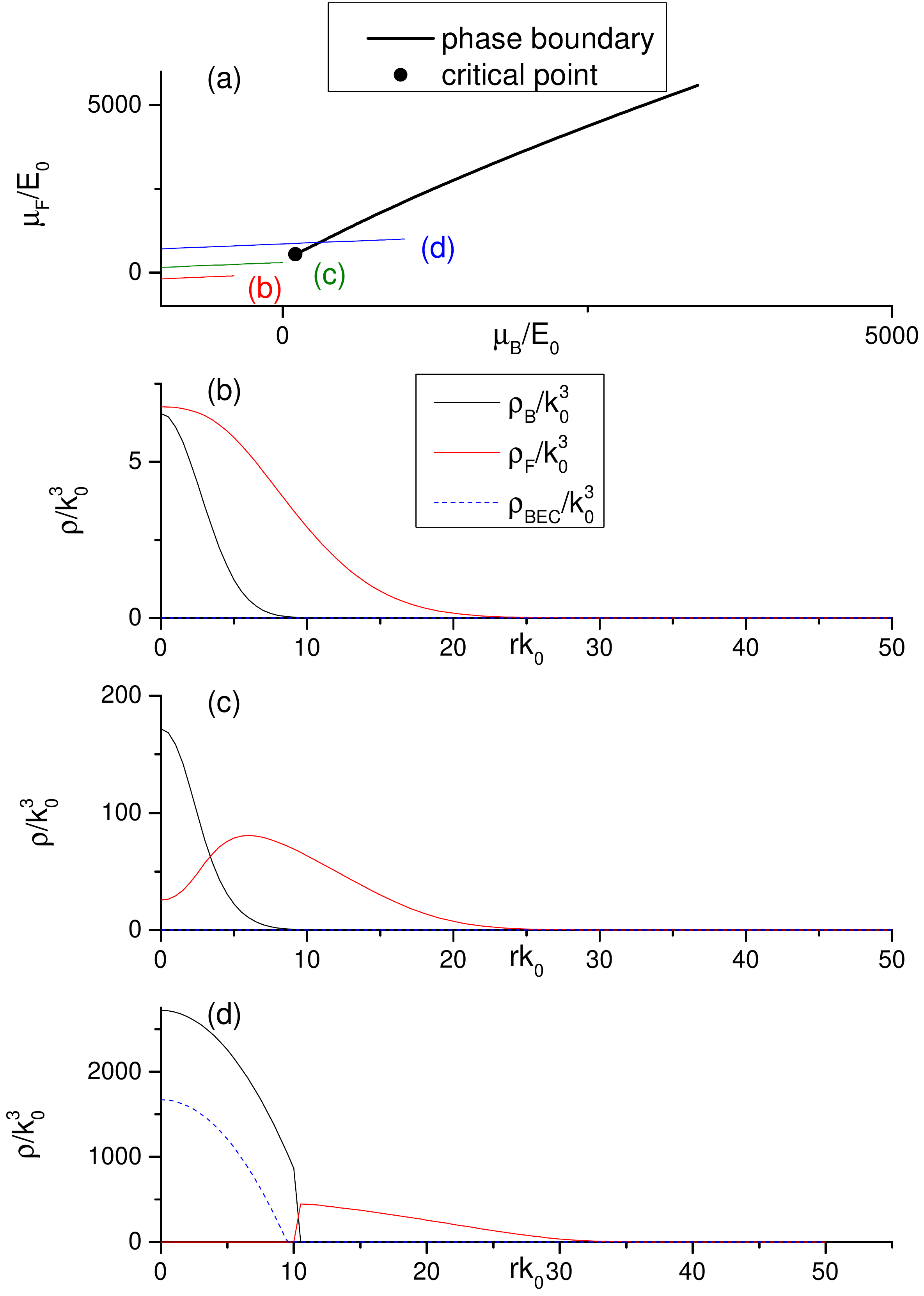}
	\caption{Harmonically trapped atomic mixtures of $^6$Li and $^{41}$K. (a) Phase separation boundary (thick line) and selected cases of the effective local chemical potentials (thin lines). The curves labeled (b), (c), (d) show the chemical potentials corresponding to the density profiles shown in panels (b), (c), (d). The convention follows Fig.~\ref{fig:phase_boundary_and_density_profile_Li6_Li7}. Here $T/T_0=100$ and $a_{BB}k_0=a_{BF}k_0=0.1$. (b) $N_B=2.09\times 10^3$ and $N_F=4.39\times10^4$. The densities of bosons and fermions both decrease monotonically from the trap center. (c) $N_B=4.21\times 10^4$ and $N_F=1.16\times 10^6$. The fermions are partially pushed away from the trap center, but the chemical potentials have not crossed the phase separation boundary. (d) $N_B=7.00\times 10^6$ and $N_F=2.07\times 10^7$. The effective local chemical potentials now cross the phase separation boundary once, and the boson-rich region discontinuously changes to the fermion-rich region.
	}
	\label{fig:phase_boundary_and_density_profile_Li6_K41}
\end{figure}

In the following we present the results of $^{6}$Li-$^{7}$Li and $^{6}$Li-$^{41}$K mixtures in harmonic traps. We emphasize that our method is general for other binary atomic Bose-Fermi mixtures, too. Figs.~\ref{fig:phase_boundary_and_density_profile_Li6_Li7}(a) and \ref{fig:phase_boundary_and_density_profile_Li6_K41}(a) show the phase separation boundary in the $\mu_B$ - $\mu_F$ parameter space with given values of $a_{BB}$, $a_{BF}$, and $T$. The phase separation boundary was constructed according to Fig.~\ref{fig:1field_muB_vs_p}, and it shows where the structural transition occurs. Since small chemical potentials imply dilute densities, there is no phase separation in the regime with small chemical potentials. Therefore, a critical point terminates lower end of the phase separation boundary.

The effective local chemical potentials of bosons and fermions in the LDA follow the $r^2$ decay. Thus, the values of $\mu_B(r)$ and $\mu_F(r)$ of a trapped mixture represent a straight line in the $\mu_B$ - $\mu_F$ parameter space with its upper-right end being the values at the trap center. Therefore, one can generate various trap density profiles by choosing different lines in the $\mu_B$ - $\mu_F$ space. For the two types of Bose-Fermi mixtures shown in Figs.~\ref{fig:phase_boundary_and_density_profile_Li6_Li7} and \ref{fig:phase_boundary_and_density_profile_Li6_K41}, there are three typical cases: The first one is when the $\mu_B(r)$ - $\mu_F(r)$ line is below the critical point of the phase separation boundary, which corresponds to the case with weak boson-fermion interactions or low densities. In this case, the two species both show monotonically decreasing density profiles. The second one is when the $\mu_B(r)$ - $\mu_F(r)$ line is close to the phase separation boundary but there is no intersection. In this case, the fermions are pushed away from the trap center and exhibit a non-monotonic trap profile. However, there is no discontinuity in the density profiles. Thus, the boson-rich region and the fermion-rich region are smoothly connected. The third case is when the $\mu_B(r)$ - $\mu_F(r)$ line intersects the phase separation boundary. Then, a genuine phase separation structure emerges, where discontinuities of the density profiles can be observed.

The density profiles of the three typical cases are illustrated in Figs.~\ref{fig:phase_boundary_and_density_profile_Li6_Li7}(b)-(d) and \ref{fig:phase_boundary_and_density_profile_Li6_K41}(b)-(d) for $^6$Li-$^7$Li and $^6$Li-$^41$K mixtures, respectively. In those plots, we fixed the temperature and interaction strengths but tune the total particle numbers. One can generate similar structures by tuning the temperature or interactions as well. We also notice it is possible the $\mu_B(r)$ - $\mu_F(r)$ line can intersect the phase separation boundary more than once. When that happens, the density profiles will exhibit sandwich structures where multiple boson-rich (or fermion-rich) regions can be found. Fig.~\ref{fig:phase_boundary_and_density_profile_Li6_Li7}(e) illustrates a possible sandwich structure with two fermion-rich regions in a harmonically trapped $^6$Li-$^7$Li mixture. For the $^6$Li-$^{41}$K mixture, it will require much larger local chemical potentials at the trap center to generate sandwich structures because of the larger mass difference. We remark that sandwich structures of binary Bose-Fermi mixtures have been discussed in Ref.~\cite{PhysRevLett.80.1804}. Here, we use a unified theoretical framework to show how various structures can emerge in a broader range of temperature and interactions accessible in experiments.

\section{Conclusion}\label{sec:conclusion}
A path-integral framework for describing binary atomic Bose-Fermi mixtures has been presented here. By integrating out the fermions in the effective action and using the large-N expansion to find the leading-order effective potential of the composite system, the nature of the mixture - phase separation transition can be visualized clearly as the free-energy curve intersects itself. For atomic gases in box potentials, the construction of phase separation is complicated by the compressibility of atomic gases, which differentiate this work from conventional liquid or solid mixtures. The phase diagrams presented here guarantees mechanical and diffusive equilibrium because the lever rule has been implemented. By using the LDA, we show typical density profiles of harmonically trapped atomic mixtures. The framework is versatile and applicable to other binary atomic mixtures. 

However, the theory has not included the higher-order effective fermion-fermion interactions due to the fermion-boson interactions, and the leading-order large-N theory underestimates the BEC transition temperature~\cite{PhysRevLett.105.240402,PhysRevA.93.033637}. The more sophisticated LOAF theory offers a better estimation, but its integration with the LDA in the strongly interacting regime remains a challenge. Therefore, while the theory presented here offers an overview of binary atomic Bose-Fermi mixtures and captures the main features, future investigations with more complicated treatments will further improve the theoretical framework.

\textit{Acknowledgement:} We thank Fred Cooper, Eddy Timmerman, Roberto Onofrio, Ming-Shien Chang, and Kelvin Wright for stimulating discussions and Bo Huang for sharing his experimental data and thoughts.

\appendix
\section{Lever Rule}
Here we explain the lever rule and how it can be applied to binary gaseous mixtures. Our method generalizes the conventional one~\cite{schroeder2013introduction,tester1997thermodynamics} applicable to liquid or solid mixtures with fixed densities. We label the two species of a binary mixture by $1$ and $2$. We use the Gibbs free energy to illustrate the lever rule because of its simple relation to the chemical potential~\cite{schroeder2013introduction}, but the derivation applies to any thermodynamic free energy by using the suitable Legendre transform.

We first divide the Gibbs free energy by the total particle number to obtain
$g\equiv G/N$,
where $N=N_1+N_2$. Since $G=N_1 \mu_1 +N_2\mu_2$~\cite{schroeder2013introduction}, $g$ becomes
\begin{eqnarray}
g=w \mu_1+(1-w)\mu_2,
\label{eq:g2}
\end{eqnarray}
where $w\equiv N_1/N$. If $g$ has a negative curvature, i.e. $\partial_w\partial_w\left(g\right)_{T,p,N}<0$, the system is unstable against phase separation. 
If the system starts with an initial mixture ratio $w_o$ and phase separates into the $\alpha$ and $\beta$ phases, the lever rule guarantees the sum of the extensive variables from the separated phases equal to the extensive variables in the initial mixture. For the present case, the lever rule is
\begin{eqnarray}
X w_\alpha +\left(1-X\right) w_\beta&&=w_o
\Rightarrow X=\frac{w_o-w_\beta}{w_\alpha-w_\beta},
\end{eqnarray}
where $X$ is the number of particles (including both species $1$ and $2$) in the $\alpha$ phase divided by the total particle number, $w_\alpha$  ($w_\beta$) is the ratio between the number of species $1$ in the $\alpha$ ($\beta$) phase and the total number of particles in phase $\alpha$ ($\beta$). In the phase separation regime, the total free energy becomes
\begin{eqnarray}
g&&
=X g(w_\alpha)+\left(1-X\right) g(w_\beta).
\end{eqnarray}

The second property of the lever rule is that it guarantees the intensive variables take the same values in the separated phases. This can be demonstrated as follows.
In equilibrium, the free energy reaches a local minimum. Thus, its derivatives should vanish:
\begin{eqnarray}
\partial_{w_\alpha}g
&&=
\left(g(w_\beta)-g(w_\alpha)\right)\frac{ w_o-w_\beta}{\left(w_\alpha-w_\beta\right)^2}\nonumber\\
&&+
\frac{ w_o-w_\beta}{w_\alpha-w_\beta}\partial_{w_\alpha}g(w_\alpha)
=0,
\end{eqnarray}
where the partial derivatives assume the variables $p,T,N$ are fixed. After simplifying the expression, the equation becomes
$\partial_{w_\alpha}g(w_\alpha)
=\frac{ g(w_\alpha)-g(w_\beta)}{w_\alpha-w_\beta}$.
A similar derivation shows the same expression for $\partial_{w_\beta}g(w_\beta)$.

Hence, we obtain
\begin{eqnarray}\label{eq:dg}
&&\partial_{w_\alpha} g(w_\alpha)=\partial_{w_\beta} g(w_\beta)=\frac{g(w_\alpha)-g(w_\beta)}{w_\alpha-w_\beta}\nonumber\\
&&\Leftrightarrow g(w_\alpha)-w_\alpha\partial_{w_\alpha} g(w_\alpha)=g(w_\beta)-w_\beta \partial_{w_\beta} g(w_\beta).
\end{eqnarray}
Next, we apply Eq.~\eqref{eq:g2} to phase $\alpha$ and phase $\beta$ separately: $g(\omega_{\alpha,\beta})=w_{\alpha,\beta} \mu_1(w_{\alpha,\beta})+(1-w_{\alpha,\beta})\mu_2(w_{\alpha,\beta})$. Then, Eq.~\eqref{eq:dg} leads to
\begin{eqnarray}
\mu_2(w_\alpha)=\mu_2(w_\beta).
\end{eqnarray}
Likewise, one can show that $\mu_1(w_\alpha)=\mu_1(w_\beta)$. Thus, diffusive equilibrium between phases $\alpha$ and $\beta$ is established. A similar derivation for the pressure will lead to mechanical equilibrium of each species between the two phases.

In this work, we use the grand potential because the chemical potentials $\mu_{B,F}$ are parts of its arguments~\cite{schroeder2013introduction,ChienPRA12}. The effective potential $V_{eff}$ is the volume density of the grand potential. When the $V_{eff}$ curve intersects itself as the chemical potentials vary, the intersection has two solutions of the extensive variable with the same values of the system parameters. The two solutions correspond to the two branches of the phase separation boundary. The lever rule then allows us to find the correct ratio of the extensive variable in each phase. The lever rule used here is
\begin{eqnarray}
	\sum_{j=\alpha,\beta}  X^j\partial_{I_{ij}} V_{eff}=\eta_{o}^i,
\end{eqnarray}
where $X^{\alpha}=X$ and $X^{\beta}=1-X$, $I_{ij}$ denotes the $i$th intensive variable in the $j$th phase, and $\eta_0^i$ denote the volume density of the $i$th extensive variable in the original mixture. Here, the index $i$ covers only the extensive variables that need to be fixed. For example, the arguments of the grand potential only have one extensive variable $V$~\cite{schroeder2013introduction}. Thus, we use $I_{ij}=-\mu_{Bj},-\mu_{Fj}$ and $\eta_o^i=\rho_{Bo},\rho_{Fo}$ to obtain the lever rules~\eqref{eq:extensive_variable_conservation_Helmholtz} for binary atomic Bose-Fermi mixtures in a box potential. Here $\mu_{Bj}$ ($\mu_{Fj}$) is the boson (fermion) chemical potential in the $j$th phase and $\rho_{Bo}$ ($\rho_{Fo}$) is the boson (fermion) density in the original unstable mixture.

\bibliographystyle{apsrev4-1}

\begin{thebibliography}{57}
	\expandafter\ifx\csname natexlab\endcsname\relax\def\natexlab#1{#1}\fi
	\expandafter\ifx\csname bibnamefont\endcsname\relax
	\def\bibnamefont#1{#1}\fi
	\expandafter\ifx\csname bibfnamefont\endcsname\relax
	\def\bibfnamefont#1{#1}\fi
	\expandafter\ifx\csname citenamefont\endcsname\relax
	\def\citenamefont#1{#1}\fi
	\expandafter\ifx\csname url\endcsname\relax
	\def\url#1{\texttt{#1}}\fi
	\expandafter\ifx\csname urlprefix\endcsname\relax\def\urlprefix{URL }\fi
	\providecommand{\bibinfo}[2]{#2}
	\providecommand{\eprint}[2][]{\url{#2}}
	
	\bibitem[{\citenamefont{Cohen}(1977)}]{10.2307/1744198}
	\bibinfo{author}{\bibfnamefont{E.~G.~D.} \bibnamefont{Cohen}},
	\bibinfo{journal}{Science} \textbf{\bibinfo{volume}{197}},
	\bibinfo{pages}{11} (\bibinfo{year}{1977}), ISSN \bibinfo{issn}{00368075,
		10959203}, \urlprefix\url{http://www.jstor.org/stable/1744198}.
	
	\bibitem[{\citenamefont{Kuerten}(1987)}]{Kuerten1987}
	\bibinfo{author}{\bibfnamefont{J.}~\bibnamefont{Kuerten}},
	\emph{\bibinfo{title}{3He-4He II mixtures : thermodynamic and hydrodynamic
			properties / door Johannes Gerardus Maria Kuerten}}
	(\bibinfo{publisher}{Eindhoven : Technische Universiteit Eindhoven},
	\bibinfo{year}{1987}), \urlprefix\url{http://dx.doi.org/10.6100/IR270164}.
	
	\bibitem[{\citenamefont{Pobell}(2013)}]{pobell2013matter}
	\bibinfo{author}{\bibfnamefont{F.}~\bibnamefont{Pobell}},
	\emph{\bibinfo{title}{Matter and Methods at Low Temperatures}}
	(\bibinfo{publisher}{Springer Berlin Heidelberg}, \bibinfo{year}{2013}), ISBN
	\bibinfo{isbn}{9783662085783},
	\urlprefix\url{https://books.google.com/books?id=H27mCAAAQBAJ}.
	
	\bibitem[{\citenamefont{Lounasmaa}(1974)}]{Lounasmaa_book}
	\bibinfo{author}{\bibfnamefont{O.~V.} \bibnamefont{Lounasmaa}},
	\emph{\bibinfo{title}{Experimental principles and methods below 1K}}
	(\bibinfo{publisher}{Academic Press}, \bibinfo{address}{Cambridge, MA, USA},
	\bibinfo{year}{1974}).
	
	\bibitem[{\citenamefont{Dalfovo et~al.}(1999)\citenamefont{Dalfovo, Giorgini,
			Pitaevskii, and Stringari}}]{RevModPhys.71.463}
	\bibinfo{author}{\bibfnamefont{F.}~\bibnamefont{Dalfovo}},
	\bibinfo{author}{\bibfnamefont{S.}~\bibnamefont{Giorgini}},
	\bibinfo{author}{\bibfnamefont{L.~P.} \bibnamefont{Pitaevskii}},
	\bibnamefont{and}
	\bibinfo{author}{\bibfnamefont{S.}~\bibnamefont{Stringari}},
	\bibinfo{journal}{Rev. Mod. Phys.} \textbf{\bibinfo{volume}{71}},
	\bibinfo{pages}{463} (\bibinfo{year}{1999}),
	\urlprefix\url{http://link.aps.org/doi/10.1103/RevModPhys.71.463}.
	
	\bibitem[{\citenamefont{Pethick and Smith}(2008)}]{pethick2008bose}
	\bibinfo{author}{\bibfnamefont{C.}~\bibnamefont{Pethick}} \bibnamefont{and}
	\bibinfo{author}{\bibfnamefont{H.}~\bibnamefont{Smith}},
	\emph{\bibinfo{title}{Bose--Einstein Condensation in Dilute Gases}}
	(\bibinfo{publisher}{Cambridge University Press}, \bibinfo{year}{2008}), ISBN
	\bibinfo{isbn}{9781139811088},
	\urlprefix\url{https://books.google.com/books?id=G8kgAwAAQBAJ}.
	
	\bibitem[{\citenamefont{Ueda}(2010)}]{ueda2010fundamentals}
	\bibinfo{author}{\bibfnamefont{M.}~\bibnamefont{Ueda}},
	\emph{\bibinfo{title}{Fundamentals and New Frontiers of Bose-Einstein
			Condensation}} (\bibinfo{publisher}{World Scientific}, \bibinfo{year}{2010}),
	ISBN \bibinfo{isbn}{9789812839596},
	\urlprefix\url{https://books.google.com/books?id=iix3\_pqy6ysC}.
	
	\bibitem[{\citenamefont{Schreck et~al.}(2001)\citenamefont{Schreck, Khaykovich,
			Corwin, Ferrari, Bourdel, Cubizolles, and Salomon}}]{PhysRevLett.87.080403}
	\bibinfo{author}{\bibfnamefont{F.}~\bibnamefont{Schreck}},
	\bibinfo{author}{\bibfnamefont{L.}~\bibnamefont{Khaykovich}},
	\bibinfo{author}{\bibfnamefont{K.~L.} \bibnamefont{Corwin}},
	\bibinfo{author}{\bibfnamefont{G.}~\bibnamefont{Ferrari}},
	\bibinfo{author}{\bibfnamefont{T.}~\bibnamefont{Bourdel}},
	\bibinfo{author}{\bibfnamefont{J.}~\bibnamefont{Cubizolles}},
	\bibnamefont{and} \bibinfo{author}{\bibfnamefont{C.}~\bibnamefont{Salomon}},
	\bibinfo{journal}{Phys. Rev. Lett.} \textbf{\bibinfo{volume}{87}},
	\bibinfo{pages}{080403} (\bibinfo{year}{2001}),
	\urlprefix\url{https://link.aps.org/doi/10.1103/PhysRevLett.87.080403}.
	
	\bibitem[{\citenamefont{Truscott et~al.}(2001)\citenamefont{Truscott, Strecker,
			McAlexander, Partridge, and Hulet}}]{Truscott2570}
	\bibinfo{author}{\bibfnamefont{A.~G.} \bibnamefont{Truscott}},
	\bibinfo{author}{\bibfnamefont{K.~E.} \bibnamefont{Strecker}},
	\bibinfo{author}{\bibfnamefont{W.~I.} \bibnamefont{McAlexander}},
	\bibinfo{author}{\bibfnamefont{G.~B.} \bibnamefont{Partridge}},
	\bibnamefont{and} \bibinfo{author}{\bibfnamefont{R.~G.} \bibnamefont{Hulet}},
	\bibinfo{journal}{Science} \textbf{\bibinfo{volume}{291}},
	\bibinfo{pages}{2570} (\bibinfo{year}{2001}), ISSN \bibinfo{issn}{0036-8075},
	\eprint{http://science.sciencemag.org/content/291/5513/2570.full.pdf},
	\urlprefix\url{http://science.sciencemag.org/content/291/5513/2570}.
	
	\bibitem[{\citenamefont{Roati et~al.}(2002)\citenamefont{Roati, Riboli,
			Modugno, and Inguscio}}]{PhysRevLett.89.150403}
	\bibinfo{author}{\bibfnamefont{G.}~\bibnamefont{Roati}},
	\bibinfo{author}{\bibfnamefont{F.}~\bibnamefont{Riboli}},
	\bibinfo{author}{\bibfnamefont{G.}~\bibnamefont{Modugno}}, \bibnamefont{and}
	\bibinfo{author}{\bibfnamefont{M.}~\bibnamefont{Inguscio}},
	\bibinfo{journal}{Phys. Rev. Lett.} \textbf{\bibinfo{volume}{89}},
	\bibinfo{pages}{150403} (\bibinfo{year}{2002}),
	\urlprefix\url{https://link.aps.org/doi/10.1103/PhysRevLett.89.150403}.
	
	\bibitem[{\citenamefont{Deh et~al.}(2008)\citenamefont{Deh, Marzok, Zimmermann,
			and Courteille}}]{PhysRevA.77.010701}
	\bibinfo{author}{\bibfnamefont{B.}~\bibnamefont{Deh}},
	\bibinfo{author}{\bibfnamefont{C.}~\bibnamefont{Marzok}},
	\bibinfo{author}{\bibfnamefont{C.}~\bibnamefont{Zimmermann}},
	\bibnamefont{and} \bibinfo{author}{\bibfnamefont{P.~W.}
		\bibnamefont{Courteille}}, \bibinfo{journal}{Phys. Rev. A}
	\textbf{\bibinfo{volume}{77}}, \bibinfo{pages}{010701}
	(\bibinfo{year}{2008}),
	\urlprefix\url{https://link.aps.org/doi/10.1103/PhysRevA.77.010701}.
	
	\bibitem[{\citenamefont{Tey et~al.}(2010)\citenamefont{Tey, Stellmer, Grimm,
			and Schreck}}]{PhysRevA.82.011608}
	\bibinfo{author}{\bibfnamefont{M.~K.} \bibnamefont{Tey}},
	\bibinfo{author}{\bibfnamefont{S.}~\bibnamefont{Stellmer}},
	\bibinfo{author}{\bibfnamefont{R.}~\bibnamefont{Grimm}}, \bibnamefont{and}
	\bibinfo{author}{\bibfnamefont{F.}~\bibnamefont{Schreck}},
	\bibinfo{journal}{Phys. Rev. A} \textbf{\bibinfo{volume}{82}},
	\bibinfo{pages}{011608} (\bibinfo{year}{2010}),
	\urlprefix\url{https://link.aps.org/doi/10.1103/PhysRevA.82.011608}.
	
	\bibitem[{\citenamefont{Yao et~al.}(2016)\citenamefont{Yao, Chen, Wu, Liu,
			Wang, Jiang, Deng, Chen, and Pan}}]{PhysRevLett.117.145301}
	\bibinfo{author}{\bibfnamefont{X.-C.} \bibnamefont{Yao}},
	\bibinfo{author}{\bibfnamefont{H.-Z.} \bibnamefont{Chen}},
	\bibinfo{author}{\bibfnamefont{Y.-P.} \bibnamefont{Wu}},
	\bibinfo{author}{\bibfnamefont{X.-P.} \bibnamefont{Liu}},
	\bibinfo{author}{\bibfnamefont{X.-Q.} \bibnamefont{Wang}},
	\bibinfo{author}{\bibfnamefont{X.}~\bibnamefont{Jiang}},
	\bibinfo{author}{\bibfnamefont{Y.}~\bibnamefont{Deng}},
	\bibinfo{author}{\bibfnamefont{Y.-A.} \bibnamefont{Chen}}, \bibnamefont{and}
	\bibinfo{author}{\bibfnamefont{J.-W.} \bibnamefont{Pan}},
	\bibinfo{journal}{Phys. Rev. Lett.} \textbf{\bibinfo{volume}{117}},
	\bibinfo{pages}{145301} (\bibinfo{year}{2016}),
	\urlprefix\url{https://link.aps.org/doi/10.1103/PhysRevLett.117.145301}.
	
	\bibitem[{\citenamefont{Lous et~al.}(2018)\citenamefont{Lous, Fritsche, Jag,
			Lehman, Kirilov, Huang, and Grimm}}]{Lous18}
	\bibinfo{author}{\bibfnamefont{R.~S.} \bibnamefont{Lous}},
	\bibinfo{author}{\bibfnamefont{I.}~\bibnamefont{Fritsche}},
	\bibinfo{author}{\bibfnamefont{M.}~\bibnamefont{Jag}},
	\bibinfo{author}{\bibfnamefont{F.}~\bibnamefont{Lehman}},
	\bibinfo{author}{\bibfnamefont{E.}~\bibnamefont{Kirilov}},
	\bibinfo{author}{\bibfnamefont{B.}~\bibnamefont{Huang}}, \bibnamefont{and}
	\bibinfo{author}{\bibfnamefont{R.}~\bibnamefont{Grimm}}
	(\bibinfo{year}{2018}), \bibinfo{note}{arXiv: 1802.01954}.
	
	\bibitem[{\citenamefont{DeSalvo et~al.}(2017)\citenamefont{DeSalvo, Patel,
			Johansen, and Chin}}]{PhysRevLett.119.233401}
	\bibinfo{author}{\bibfnamefont{B.~J.} \bibnamefont{DeSalvo}},
	\bibinfo{author}{\bibfnamefont{K.}~\bibnamefont{Patel}},
	\bibinfo{author}{\bibfnamefont{J.}~\bibnamefont{Johansen}}, \bibnamefont{and}
	\bibinfo{author}{\bibfnamefont{C.}~\bibnamefont{Chin}},
	\bibinfo{journal}{Phys. Rev. Lett.} \textbf{\bibinfo{volume}{119}},
	\bibinfo{pages}{233401} (\bibinfo{year}{2017}),
	\urlprefix\url{https://link.aps.org/doi/10.1103/PhysRevLett.119.233401}.
	
	\bibitem[{\citenamefont{Onofrio}(2016)}]{1063-7869-59-11-1129}
	\bibinfo{author}{\bibfnamefont{R.}~\bibnamefont{Onofrio}},
	\bibinfo{journal}{Physics-Uspekhi} \textbf{\bibinfo{volume}{59}},
	\bibinfo{pages}{1129} (\bibinfo{year}{2016}),
	\urlprefix\url{http://stacks.iop.org/1063-7869/59/i=11/a=1129}.
	
	\bibitem[{\citenamefont{Presilla and Onofrio}(2003)}]{Presilla03}
	\bibinfo{author}{\bibfnamefont{C.}~\bibnamefont{Presilla}} \bibnamefont{and}
	\bibinfo{author}{\bibfnamefont{R.}~\bibnamefont{Onofrio}},
	\bibinfo{journal}{Phys. Rev. Lett.} \textbf{\bibinfo{volume}{90}},
	\bibinfo{pages}{030404} (\bibinfo{year}{2003}).
	
	\bibitem[{\citenamefont{Donley et~al.}(2001)\citenamefont{Donley, Claussen,
			Cornish, Roberts, Cornell, and Wieman}}]{Donley2001}
	\bibinfo{author}{\bibfnamefont{E.~A.} \bibnamefont{Donley}},
	\bibinfo{author}{\bibfnamefont{N.~R.} \bibnamefont{Claussen}},
	\bibinfo{author}{\bibfnamefont{S.~L.} \bibnamefont{Cornish}},
	\bibinfo{author}{\bibfnamefont{J.~L.} \bibnamefont{Roberts}},
	\bibinfo{author}{\bibfnamefont{E.~A.} \bibnamefont{Cornell}},
	\bibnamefont{and} \bibinfo{author}{\bibfnamefont{C.~E.}
		\bibnamefont{Wieman}}, \bibinfo{journal}{Nature}
	\textbf{\bibinfo{volume}{412}}, \bibinfo{pages}{295} (\bibinfo{year}{2001}),
	\urlprefix\url{http://dx.doi.org/10.1038/35085500}.
	
	\bibitem[{\citenamefont{M\o{}lmer}(1998)}]{PhysRevLett.80.1804}
	\bibinfo{author}{\bibfnamefont{K.}~\bibnamefont{M\o{}lmer}},
	\bibinfo{journal}{Phys. Rev. Lett.} \textbf{\bibinfo{volume}{80}},
	\bibinfo{pages}{1804} (\bibinfo{year}{1998}),
	\urlprefix\url{https://link.aps.org/doi/10.1103/PhysRevLett.80.1804}.
	
	\bibitem[{\citenamefont{Nygaard and M\o{}lmer}(1999)}]{PhysRevA.59.2974}
	\bibinfo{author}{\bibfnamefont{N.}~\bibnamefont{Nygaard}} \bibnamefont{and}
	\bibinfo{author}{\bibfnamefont{K.}~\bibnamefont{M\o{}lmer}},
	\bibinfo{journal}{Phys. Rev. A} \textbf{\bibinfo{volume}{59}},
	\bibinfo{pages}{2974} (\bibinfo{year}{1999}),
	\urlprefix\url{https://link.aps.org/doi/10.1103/PhysRevA.59.2974}.
	
	\bibitem[{\citenamefont{Viverit et~al.}(2000)\citenamefont{Viverit, Pethick,
			and Smith}}]{PhysRevA.61.053605}
	\bibinfo{author}{\bibfnamefont{L.}~\bibnamefont{Viverit}},
	\bibinfo{author}{\bibfnamefont{C.~J.} \bibnamefont{Pethick}},
	\bibnamefont{and} \bibinfo{author}{\bibfnamefont{H.}~\bibnamefont{Smith}},
	\bibinfo{journal}{Phys. Rev. A} \textbf{\bibinfo{volume}{61}},
	\bibinfo{pages}{053605} (\bibinfo{year}{2000}),
	\urlprefix\url{https://link.aps.org/doi/10.1103/PhysRevA.61.053605}.
	
	\bibitem[{\citenamefont{Malatsetxebarria
			et~al.}(2013)\citenamefont{Malatsetxebarria, Marchetti, and
			Cazalilla}}]{PhysRevA.88.033604}
	\bibinfo{author}{\bibfnamefont{E.}~\bibnamefont{Malatsetxebarria}},
	\bibinfo{author}{\bibfnamefont{F.~M.} \bibnamefont{Marchetti}},
	\bibnamefont{and} \bibinfo{author}{\bibfnamefont{M.~A.}
		\bibnamefont{Cazalilla}}, \bibinfo{journal}{Phys. Rev. A}
	\textbf{\bibinfo{volume}{88}}, \bibinfo{pages}{033604}
	(\bibinfo{year}{2013}),
	\urlprefix\url{https://link.aps.org/doi/10.1103/PhysRevA.88.033604}.
	
	\bibitem[{\citenamefont{Amoruso et~al.}(1998)\citenamefont{Amoruso, Minguzzi,
			Stringari, Tosi, and Vichi}}]{Amoruso1998}
	\bibinfo{author}{\bibfnamefont{M.}~\bibnamefont{Amoruso}},
	\bibinfo{author}{\bibfnamefont{A.}~\bibnamefont{Minguzzi}},
	\bibinfo{author}{\bibfnamefont{S.}~\bibnamefont{Stringari}},
	\bibinfo{author}{\bibfnamefont{M.}~\bibnamefont{Tosi}}, \bibnamefont{and}
	\bibinfo{author}{\bibfnamefont{L.}~\bibnamefont{Vichi}},
	\bibinfo{journal}{Eur. Phys. J. D} \textbf{\bibinfo{volume}{4}},
	\bibinfo{pages}{261} (\bibinfo{year}{1998}), ISSN \bibinfo{issn}{1434-6079},
	\urlprefix\url{https://doi.org/10.1007/s100530050207}.
	
	\bibitem[{\citenamefont{Akdeniz et~al.}(2002)\citenamefont{Akdeniz, Vignolo,
			Minguzzi, and Tosi}}]{0953-4075-35-4-102}
	\bibinfo{author}{\bibfnamefont{Z.}~\bibnamefont{Akdeniz}},
	\bibinfo{author}{\bibfnamefont{P.}~\bibnamefont{Vignolo}},
	\bibinfo{author}{\bibfnamefont{A.}~\bibnamefont{Minguzzi}}, \bibnamefont{and}
	\bibinfo{author}{\bibfnamefont{M.~P.} \bibnamefont{Tosi}},
	\bibinfo{journal}{J. Phys. B: At. Mol. Opt. Phys.}
	\textbf{\bibinfo{volume}{35}}, \bibinfo{pages}{L105} (\bibinfo{year}{2002}),
	\urlprefix\url{http://stacks.iop.org/0953-4075/35/i=4/a=102}.
	
	\bibitem[{\citenamefont{Snoek et~al.}(2005)\citenamefont{Snoek, Haque,
			Vandoren, and Stoof}}]{Snoek05}
	\bibinfo{author}{\bibfnamefont{M.}~\bibnamefont{Snoek}},
	\bibinfo{author}{\bibfnamefont{M.}~\bibnamefont{Haque}},
	\bibinfo{author}{\bibfnamefont{S.}~\bibnamefont{Vandoren}}, \bibnamefont{and}
	\bibinfo{author}{\bibfnamefont{H.~T.~C.} \bibnamefont{Stoof}},
	\bibinfo{journal}{Phys. Rev. Lett.} \textbf{\bibinfo{volume}{95}},
	\bibinfo{pages}{250401} (\bibinfo{year}{2005}).
	
	\bibitem[{\citenamefont{Yu and Yang}(2008)}]{Yu08}
	\bibinfo{author}{\bibfnamefont{Y.}~\bibnamefont{Yu}} \bibnamefont{and}
	\bibinfo{author}{\bibfnamefont{K.}~\bibnamefont{Yang}},
	\bibinfo{journal}{Phys. Rev. Lett.} \textbf{\bibinfo{volume}{100}},
	\bibinfo{pages}{090404} (\bibinfo{year}{2008}).
	
	\bibitem[{\citenamefont{Maeda et~al.}(2009)\citenamefont{Maeda, Baym, and
			Hatsuda}}]{Maeda09}
	\bibinfo{author}{\bibfnamefont{K.}~\bibnamefont{Maeda}},
	\bibinfo{author}{\bibfnamefont{G.}~\bibnamefont{Baym}}, \bibnamefont{and}
	\bibinfo{author}{\bibfnamefont{T.}~\bibnamefont{Hatsuda}},
	\bibinfo{journal}{Phys. Rev. Lett.} \textbf{\bibinfo{volume}{103}},
	\bibinfo{pages}{085301} (\bibinfo{year}{2009}).
	
	\bibitem[{\citenamefont{Cooper et~al.}(2010)\citenamefont{Cooper, Chien,
			Mihaila, Dawson, and Timmermans}}]{PhysRevLett.105.240402}
	\bibinfo{author}{\bibfnamefont{F.}~\bibnamefont{Cooper}},
	\bibinfo{author}{\bibfnamefont{C.-C.} \bibnamefont{Chien}},
	\bibinfo{author}{\bibfnamefont{B.}~\bibnamefont{Mihaila}},
	\bibinfo{author}{\bibfnamefont{J.~F.} \bibnamefont{Dawson}},
	\bibnamefont{and}
	\bibinfo{author}{\bibfnamefont{E.}~\bibnamefont{Timmermans}},
	\bibinfo{journal}{Phys. Rev. Lett.} \textbf{\bibinfo{volume}{105}},
	\bibinfo{pages}{240402} (\bibinfo{year}{2010}),
	\urlprefix\url{http://link.aps.org/doi/10.1103/PhysRevLett.105.240402}.
	
	\bibitem[{\citenamefont{Chien et~al.}(2012)\citenamefont{Chien, Cooper, and
			Timmermans}}]{ChienPRA12}
	\bibinfo{author}{\bibfnamefont{C.-C.} \bibnamefont{Chien}},
	\bibinfo{author}{\bibfnamefont{F.}~\bibnamefont{Cooper}}, \bibnamefont{and}
	\bibinfo{author}{\bibfnamefont{E.}~\bibnamefont{Timmermans}},
	\bibinfo{journal}{Phys. Rev. A} \textbf{\bibinfo{volume}{86}},
	\bibinfo{pages}{023634} (\bibinfo{year}{2012}),
	\urlprefix\url{http://link.aps.org/doi/10.1103/PhysRevA.86.023634}.
	
	\bibitem[{\citenamefont{Cooper et~al.}(2011)\citenamefont{Cooper, Mihaila,
			Dawson, Chien, and Timmermans}}]{PhysRevA.83.053622}
	\bibinfo{author}{\bibfnamefont{F.}~\bibnamefont{Cooper}},
	\bibinfo{author}{\bibfnamefont{B.}~\bibnamefont{Mihaila}},
	\bibinfo{author}{\bibfnamefont{J.~F.} \bibnamefont{Dawson}},
	\bibinfo{author}{\bibfnamefont{C.-C.} \bibnamefont{Chien}}, \bibnamefont{and}
	\bibinfo{author}{\bibfnamefont{E.}~\bibnamefont{Timmermans}},
	\bibinfo{journal}{Phys. Rev. A} \textbf{\bibinfo{volume}{83}},
	\bibinfo{pages}{053622} (\bibinfo{year}{2011}),
	\urlprefix\url{http://link.aps.org/doi/10.1103/PhysRevA.83.053622}.
	
	\bibitem[{\citenamefont{Kim and Chien}(2016)}]{PhysRevA.93.033637}
	\bibinfo{author}{\bibfnamefont{T.}~\bibnamefont{Kim}} \bibnamefont{and}
	\bibinfo{author}{\bibfnamefont{C.-C.} \bibnamefont{Chien}},
	\bibinfo{journal}{Phys. Rev. A} \textbf{\bibinfo{volume}{93}},
	\bibinfo{pages}{033637} (\bibinfo{year}{2016}),
	\urlprefix\url{https://link.aps.org/doi/10.1103/PhysRevA.93.033637}.
	
	\bibitem[{\citenamefont{Schroeder}(2013)}]{schroeder2013introduction}
	\bibinfo{author}{\bibfnamefont{D.}~\bibnamefont{Schroeder}},
	\emph{\bibinfo{title}{An Introduction to Thermal Physics}}, Always learning
	(\bibinfo{publisher}{Pearson Education, Limited}, \bibinfo{year}{2013}), ISBN
	\bibinfo{isbn}{9781292026213},
	\urlprefix\url{https://books.google.com/books?id=XjECngEACAAJ}.
	
	\bibitem[{\citenamefont{Ikemachi et~al.}(2017)\citenamefont{Ikemachi, Ito,
			Aratake, Chen, Koashi, Kuwata-Gonokami, and
			Horikoshi}}]{0953-4075-50-1-01LT01}
	\bibinfo{author}{\bibfnamefont{T.}~\bibnamefont{Ikemachi}},
	\bibinfo{author}{\bibfnamefont{A.}~\bibnamefont{Ito}},
	\bibinfo{author}{\bibfnamefont{Y.}~\bibnamefont{Aratake}},
	\bibinfo{author}{\bibfnamefont{Y.}~\bibnamefont{Chen}},
	\bibinfo{author}{\bibfnamefont{M.}~\bibnamefont{Koashi}},
	\bibinfo{author}{\bibfnamefont{M.}~\bibnamefont{Kuwata-Gonokami}},
	\bibnamefont{and}
	\bibinfo{author}{\bibfnamefont{M.}~\bibnamefont{Horikoshi}},
	\bibinfo{journal}{J. Phys. B: At. Mol. Opt. Phys.}
	\textbf{\bibinfo{volume}{50}}, \bibinfo{pages}{01LT01}
	(\bibinfo{year}{2017}),
	\urlprefix\url{http://stacks.iop.org/0953-4075/50/i=1/a=01LT01}.
	
	\bibitem[{\citenamefont{Ferrier-Barbut
			et~al.}(2014)\citenamefont{Ferrier-Barbut, Delehaye, Laurent, Grier, Pierce,
			Rem, Chevy, and Salomon}}]{Ferrier-Barbut1035}
	\bibinfo{author}{\bibfnamefont{I.}~\bibnamefont{Ferrier-Barbut}},
	\bibinfo{author}{\bibfnamefont{M.}~\bibnamefont{Delehaye}},
	\bibinfo{author}{\bibfnamefont{S.}~\bibnamefont{Laurent}},
	\bibinfo{author}{\bibfnamefont{A.~T.} \bibnamefont{Grier}},
	\bibinfo{author}{\bibfnamefont{M.}~\bibnamefont{Pierce}},
	\bibinfo{author}{\bibfnamefont{B.~S.} \bibnamefont{Rem}},
	\bibinfo{author}{\bibfnamefont{F.}~\bibnamefont{Chevy}}, \bibnamefont{and}
	\bibinfo{author}{\bibfnamefont{C.}~\bibnamefont{Salomon}},
	\bibinfo{journal}{Science} \textbf{\bibinfo{volume}{345}},
	\bibinfo{pages}{1035} (\bibinfo{year}{2014}), ISSN \bibinfo{issn}{0036-8075},
	\eprint{http://science.sciencemag.org/content/345/6200/1035.full.pdf},
	\urlprefix\url{http://science.sciencemag.org/content/345/6200/1035}.
	
	\bibitem[{\citenamefont{Roy et~al.}(2017)\citenamefont{Roy, Green, Bowler, and
			Gupta}}]{PhysRevLett.118.055301}
	\bibinfo{author}{\bibfnamefont{R.}~\bibnamefont{Roy}},
	\bibinfo{author}{\bibfnamefont{A.}~\bibnamefont{Green}},
	\bibinfo{author}{\bibfnamefont{R.}~\bibnamefont{Bowler}}, \bibnamefont{and}
	\bibinfo{author}{\bibfnamefont{S.}~\bibnamefont{Gupta}},
	\bibinfo{journal}{Phys. Rev. Lett.} \textbf{\bibinfo{volume}{118}},
	\bibinfo{pages}{055301} (\bibinfo{year}{2017}),
	\urlprefix\url{https://link.aps.org/doi/10.1103/PhysRevLett.118.055301}.
	
	\bibitem[{\citenamefont{Hall et~al.}(1998)\citenamefont{Hall, Matthews, Ensher,
			Wieman, and Cornell}}]{PhysRevLett.81.1539}
	\bibinfo{author}{\bibfnamefont{D.~S.} \bibnamefont{Hall}},
	\bibinfo{author}{\bibfnamefont{M.~R.} \bibnamefont{Matthews}},
	\bibinfo{author}{\bibfnamefont{J.~R.} \bibnamefont{Ensher}},
	\bibinfo{author}{\bibfnamefont{C.~E.} \bibnamefont{Wieman}},
	\bibnamefont{and} \bibinfo{author}{\bibfnamefont{E.~A.}
		\bibnamefont{Cornell}}, \bibinfo{journal}{Phys. Rev. Lett.}
	\textbf{\bibinfo{volume}{81}}, \bibinfo{pages}{1539} (\bibinfo{year}{1998}),
	\urlprefix\url{https://link.aps.org/doi/10.1103/PhysRevLett.81.1539}.
	
	\bibitem[{\citenamefont{Cui and Ho}(2013)}]{PhysRevLett.110.165302}
	\bibinfo{author}{\bibfnamefont{X.}~\bibnamefont{Cui}} \bibnamefont{and}
	\bibinfo{author}{\bibfnamefont{T.-L.} \bibnamefont{Ho}},
	\bibinfo{journal}{Phys. Rev. Lett.} \textbf{\bibinfo{volume}{110}},
	\bibinfo{pages}{165302} (\bibinfo{year}{2013}),
	\urlprefix\url{https://link.aps.org/doi/10.1103/PhysRevLett.110.165302}.
	
	\bibitem[{\citenamefont{Fratini and Pilati}(2014)}]{PhysRevA.90.023605}
	\bibinfo{author}{\bibfnamefont{E.}~\bibnamefont{Fratini}} \bibnamefont{and}
	\bibinfo{author}{\bibfnamefont{S.}~\bibnamefont{Pilati}},
	\bibinfo{journal}{Phys. Rev. A} \textbf{\bibinfo{volume}{90}},
	\bibinfo{pages}{023605} (\bibinfo{year}{2014}),
	\urlprefix\url{https://link.aps.org/doi/10.1103/PhysRevA.90.023605}.
	
	\bibitem[{\citenamefont{Pecak et~al.}(2016)\citenamefont{Pecak, Gajda, and
			Sowinski}}]{1367-2630-18-1-013030}
	\bibinfo{author}{\bibfnamefont{D.}~\bibnamefont{Pecak}},
	\bibinfo{author}{\bibfnamefont{M.}~\bibnamefont{Gajda}}, \bibnamefont{and}
	\bibinfo{author}{\bibfnamefont{T.}~\bibnamefont{Sowinski}},
	\bibinfo{journal}{New J. Phys.} \textbf{\bibinfo{volume}{18}},
	\bibinfo{pages}{013030} (\bibinfo{year}{2016}),
	\urlprefix\url{http://stacks.iop.org/1367-2630/18/i=1/a=013030}.
	
	\bibitem[{\citenamefont{Jo et~al.}(2009)\citenamefont{Jo, Lee, Choi,
			Christensen, Kim, Thywissen, Pritchard, and Ketterle}}]{Jo1521}
	\bibinfo{author}{\bibfnamefont{G.-B.} \bibnamefont{Jo}},
	\bibinfo{author}{\bibfnamefont{Y.-R.} \bibnamefont{Lee}},
	\bibinfo{author}{\bibfnamefont{J.-H.} \bibnamefont{Choi}},
	\bibinfo{author}{\bibfnamefont{C.~A.} \bibnamefont{Christensen}},
	\bibinfo{author}{\bibfnamefont{T.~H.} \bibnamefont{Kim}},
	\bibinfo{author}{\bibfnamefont{J.~H.} \bibnamefont{Thywissen}},
	\bibinfo{author}{\bibfnamefont{D.~E.} \bibnamefont{Pritchard}},
	\bibnamefont{and} \bibinfo{author}{\bibfnamefont{W.}~\bibnamefont{Ketterle}},
	\bibinfo{journal}{Science} \textbf{\bibinfo{volume}{325}},
	\bibinfo{pages}{1521} (\bibinfo{year}{2009}), ISSN \bibinfo{issn}{0036-8075},
	\eprint{http://science.sciencemag.org/content/325/5947/1521.full.pdf},
	\urlprefix\url{http://science.sciencemag.org/content/325/5947/1521}.
	
	\bibitem[{\citenamefont{Zee}(2010)}]{zee2010quantum}
	\bibinfo{author}{\bibfnamefont{A.}~\bibnamefont{Zee}},
	\emph{\bibinfo{title}{Quantum Field Theory in a Nutshell: (Second Edition)}},
	In a Nutshell (\bibinfo{publisher}{Princeton University Press},
	\bibinfo{year}{2010}), ISBN \bibinfo{isbn}{9781400835324},
	\urlprefix\url{https://books.google.com/books?id=n8Mmbjtco78C}.
	
	\bibitem[{\citenamefont{Schakel}(2008)}]{Schakel_book}
	\bibinfo{author}{\bibfnamefont{A.~M.~J.} \bibnamefont{Schakel}},
	\emph{\bibinfo{title}{Boulevard of Broken Symmetries: Effective Field
			Theories of Condensed Matter}} (\bibinfo{publisher}{World Scientific
		Publishing}, \bibinfo{address}{Singapore}, \bibinfo{year}{2008}).
	
	\bibitem[{\citenamefont{Nair}(2006)}]{nair2006quantum}
	\bibinfo{author}{\bibfnamefont{V.}~\bibnamefont{Nair}},
	\emph{\bibinfo{title}{Quantum Field Theory: A Modern Perspective}}, Graduate
	Texts in Contemporary Physics (\bibinfo{publisher}{Springer New York},
	\bibinfo{year}{2006}), ISBN \bibinfo{isbn}{9780387250984},
	\urlprefix\url{https://books.google.com/books?id=rcr0fxDf8jsC}.
	
	\bibitem[{\citenamefont{Kittel and Kroemer}(1980)}]{kittel1980thermal}
	\bibinfo{author}{\bibfnamefont{C.}~\bibnamefont{Kittel}} \bibnamefont{and}
	\bibinfo{author}{\bibfnamefont{H.}~\bibnamefont{Kroemer}},
	\emph{\bibinfo{title}{Thermal Physics}} (\bibinfo{publisher}{W. H. Freeman},
	\bibinfo{year}{1980}), ISBN \bibinfo{isbn}{9780716710882},
	\urlprefix\url{https://books.google.com/books?id=X4sfAQAAMAAJ}.
	
	\bibitem[{\citenamefont{Reichl}(2009)}]{reichl2009modern}
	\bibinfo{author}{\bibfnamefont{L.}~\bibnamefont{Reichl}},
	\emph{\bibinfo{title}{A Modern Course in Statistical Physics}}, Physics
	textbook (\bibinfo{publisher}{Wiley}, \bibinfo{year}{2009}), ISBN
	\bibinfo{isbn}{9783527407828},
	\urlprefix\url{https://books.google.com/books?id=H\_4qxPazAYEC}.
	
	\bibitem[{\citenamefont{Chaikin and Lubensky}(1995)}]{chaikin1995principles}
	\bibinfo{author}{\bibfnamefont{P.}~\bibnamefont{Chaikin}} \bibnamefont{and}
	\bibinfo{author}{\bibfnamefont{T.}~\bibnamefont{Lubensky}},
	\emph{\bibinfo{title}{Principles of Condensed Matter Physics}}
	(\bibinfo{publisher}{Cambridge University Press}, \bibinfo{year}{1995}), ISBN
	\bibinfo{isbn}{9781139648615},
	\urlprefix\url{https://books.google.com/books?id=Bw\_ooQEACAAJ}.
	
	\bibitem[{\citenamefont{Tester and Modell}(1997)}]{tester1997thermodynamics}
	\bibinfo{author}{\bibfnamefont{J.}~\bibnamefont{Tester}} \bibnamefont{and}
	\bibinfo{author}{\bibfnamefont{M.}~\bibnamefont{Modell}},
	\emph{\bibinfo{title}{Thermodynamics and Its Applications}}, PHYSICAL AND
	CHEMICAL ENGINEERING SCIENCES (\bibinfo{publisher}{Prentice Hall PTR},
	\bibinfo{year}{1997}), ISBN \bibinfo{isbn}{9780139153563},
	\urlprefix\url{https://books.google.com/books?id=DiVhQgAACAAJ}.
	
	\bibitem[{\citenamefont{Gaunt et~al.}(2013)\citenamefont{Gaunt, Schmidutz,
			Gotlibovych, Smith, and Hadzibabic}}]{Gaunt2013}
	\bibinfo{author}{\bibfnamefont{A.~L.} \bibnamefont{Gaunt}},
	\bibinfo{author}{\bibfnamefont{T.~F.} \bibnamefont{Schmidutz}},
	\bibinfo{author}{\bibfnamefont{I.}~\bibnamefont{Gotlibovych}},
	\bibinfo{author}{\bibfnamefont{R.~P.} \bibnamefont{Smith}}, \bibnamefont{and}
	\bibinfo{author}{\bibfnamefont{Z.}~\bibnamefont{Hadzibabic}},
	\bibinfo{journal}{Phys. Rev. Lett.} \textbf{\bibinfo{volume}{110}},
	\bibinfo{pages}{200406} (\bibinfo{year}{2013}),
	\urlprefix\url{http://link.aps.org/doi/10.1103/PhysRevLett.110.200406}.
	
	\bibitem[{\citenamefont{Schmidutz et~al.}(2014)\citenamefont{Schmidutz,
			Gotlibovych, Gaunt, Smith, Navon, and Hadzibabic}}]{PhysRevLett.112.040403}
	\bibinfo{author}{\bibfnamefont{T.~F.} \bibnamefont{Schmidutz}},
	\bibinfo{author}{\bibfnamefont{I.}~\bibnamefont{Gotlibovych}},
	\bibinfo{author}{\bibfnamefont{A.~L.} \bibnamefont{Gaunt}},
	\bibinfo{author}{\bibfnamefont{R.~P.} \bibnamefont{Smith}},
	\bibinfo{author}{\bibfnamefont{N.}~\bibnamefont{Navon}}, \bibnamefont{and}
	\bibinfo{author}{\bibfnamefont{Z.}~\bibnamefont{Hadzibabic}},
	\bibinfo{journal}{Phys. Rev. Lett.} \textbf{\bibinfo{volume}{112}},
	\bibinfo{pages}{040403} (\bibinfo{year}{2014}),
	\urlprefix\url{https://link.aps.org/doi/10.1103/PhysRevLett.112.040403}.
	
	\bibitem[{\citenamefont{Lopes et~al.}(2017)\citenamefont{Lopes, Eigen, Navon,
			Cl\'ement, Smith, and Hadzibabic}}]{PhysRevLett.119.190404}
	\bibinfo{author}{\bibfnamefont{R.}~\bibnamefont{Lopes}},
	\bibinfo{author}{\bibfnamefont{C.}~\bibnamefont{Eigen}},
	\bibinfo{author}{\bibfnamefont{N.}~\bibnamefont{Navon}},
	\bibinfo{author}{\bibfnamefont{D.}~\bibnamefont{Cl\'ement}},
	\bibinfo{author}{\bibfnamefont{R.~P.} \bibnamefont{Smith}}, \bibnamefont{and}
	\bibinfo{author}{\bibfnamefont{Z.}~\bibnamefont{Hadzibabic}},
	\bibinfo{journal}{Phys. Rev. Lett.} \textbf{\bibinfo{volume}{119}},
	\bibinfo{pages}{190404} (\bibinfo{year}{2017}),
	\urlprefix\url{https://link.aps.org/doi/10.1103/PhysRevLett.119.190404}.
	
	\bibitem[{\citenamefont{Pitaevskii and Stringari}(2003)}]{pitaevskii2003bose}
	\bibinfo{author}{\bibfnamefont{L.}~\bibnamefont{Pitaevskii}} \bibnamefont{and}
	\bibinfo{author}{\bibfnamefont{S.}~\bibnamefont{Stringari}},
	\emph{\bibinfo{title}{Bose-Einstein Condensation}}, International Series of
	Monographs on Physics (\bibinfo{publisher}{Clarendon Press},
	\bibinfo{year}{2003}), ISBN \bibinfo{isbn}{9780198507192},
	\urlprefix\url{https://books.google.com/books?id=rIobbOxC4j4C}.
	
	\bibitem[{\citenamefont{Campbell and Campbell}(1937)}]{doi:10.1021/ja01291a001}
	\bibinfo{author}{\bibfnamefont{A.~N.} \bibnamefont{Campbell}} \bibnamefont{and}
	\bibinfo{author}{\bibfnamefont{A.~J.~R.} \bibnamefont{Campbell}},
	\bibinfo{journal}{J. Am. Chem. Soc.} \textbf{\bibinfo{volume}{59}},
	\bibinfo{pages}{2481} (\bibinfo{year}{1937}),
	\eprint{http://dx.doi.org/10.1021/ja01291a001},
	\urlprefix\url{http://dx.doi.org/10.1021/ja01291a001}.
	
	\bibitem[{\citenamefont{Marchetti et~al.}(2008)\citenamefont{Marchetti, Mathy,
			Huse, and Parish}}]{PhysRevB.78.134517}
	\bibinfo{author}{\bibfnamefont{F.~M.} \bibnamefont{Marchetti}},
	\bibinfo{author}{\bibfnamefont{C.~J.~M.} \bibnamefont{Mathy}},
	\bibinfo{author}{\bibfnamefont{D.~A.} \bibnamefont{Huse}}, \bibnamefont{and}
	\bibinfo{author}{\bibfnamefont{M.~M.} \bibnamefont{Parish}},
	\bibinfo{journal}{Phys. Rev. B} \textbf{\bibinfo{volume}{78}},
	\bibinfo{pages}{134517} (\bibinfo{year}{2008}),
	\urlprefix\url{https://link.aps.org/doi/10.1103/PhysRevB.78.134517}.
	
	\bibitem[{\citenamefont{van Haeringen et~al.}(1979)\citenamefont{van Haeringen,
			Staas, and Geurst}}]{BERNTSEN1979107}
	\bibinfo{author}{\bibfnamefont{W.}~\bibnamefont{van Haeringen}},
	\bibinfo{author}{\bibfnamefont{F.~A.} \bibnamefont{Staas}}, \bibnamefont{and}
	\bibinfo{author}{\bibfnamefont{J.~A.} \bibnamefont{Geurst}},
	\bibinfo{journal}{Philips J. Res.} \textbf{\bibinfo{volume}{34}},
	\bibinfo{pages}{107 } (\bibinfo{year}{1979}), ISSN \bibinfo{issn}{0165-5817},
	\urlprefix\url{http://www.extra.research.philips.com/hera/people/aarts/_Philips%20Bound%20Archive/PJR/PJR-33_34-1978_79_C-107.pdf}.
		
		\bibitem[{\citenamefont{Ebner and Edwards}(1971)}]{EBNER197177}
		\bibinfo{author}{\bibfnamefont{C.}~\bibnamefont{Ebner}} \bibnamefont{and}
		\bibinfo{author}{\bibfnamefont{D.}~\bibnamefont{Edwards}},
		\bibinfo{journal}{Phys. Rep.} \textbf{\bibinfo{volume}{2}},
		\bibinfo{pages}{77 } (\bibinfo{year}{1971}), ISSN \bibinfo{issn}{0370-1573},
		\urlprefix\url{http://www.sciencedirect.com/science/article/pii/0370157371900032}.
		
		\bibitem[{\citenamefont{Lous et~al.}(2017)\citenamefont{Lous, Fritsche, Jag,
				Huang, and Grimm}}]{PhysRevA.95.053627}
		\bibinfo{author}{\bibfnamefont{R.~S.} \bibnamefont{Lous}},
		\bibinfo{author}{\bibfnamefont{I.}~\bibnamefont{Fritsche}},
		\bibinfo{author}{\bibfnamefont{M.}~\bibnamefont{Jag}},
		\bibinfo{author}{\bibfnamefont{B.}~\bibnamefont{Huang}}, \bibnamefont{and}
		\bibinfo{author}{\bibfnamefont{R.}~\bibnamefont{Grimm}},
		\bibinfo{journal}{Phys. Rev. A} \textbf{\bibinfo{volume}{95}},
		\bibinfo{pages}{053627} (\bibinfo{year}{2017}),
		\urlprefix\url{https://link.aps.org/doi/10.1103/PhysRevA.95.053627}.
		
		\bibitem[{\citenamefont{Pathria and Beale}(2011)}]{pathria2011statistical}
		\bibinfo{author}{\bibfnamefont{R.}~\bibnamefont{Pathria}} \bibnamefont{and}
		\bibinfo{author}{\bibfnamefont{P.}~\bibnamefont{Beale}},
		\emph{\bibinfo{title}{Statistical Mechanics}} (\bibinfo{publisher}{Elsevier
			Science}, \bibinfo{year}{2011}), ISBN \bibinfo{isbn}{9780123821898},
		\urlprefix\url{https://books.google.com/books?id=KdbJJAXQ-RsC}.
		
	\end{thebibliography}

\end{document}